\documentclass[11pt,a4paper]{article}
\pdfoutput=1
\usepackage{jheppub}
\usepackage{amsfonts}
\usepackage{bbm}
\usepackage{graphicx}
\usepackage{caption}
\usepackage{subcaption}
\usepackage{bigints}
\usepackage[normalem]{ulem}
\usepackage{slashed}
\usepackage{simpler-wick}
\usepackage[export]{adjustbox}
\usepackage{tikz-feynman}
\usepackage{simpler-wick}

\def\c{\gamma} \def\g{\gamma}

\def\e{\epsilon}

\def\m{\mu}

\newcommand{\RN}[1]{%
  \textup{\uppercase\expandafter{\romannumeral#1}}%
}

\newcommand{\nn}{\nonumber}

\def\e{\epsilon}

\def\p{\partial}

\def\bea{\begin{eqnarray}}
\def\eea{\end{eqnarray}}
\def\be{\begin{equation}}
\def\ee{\end{equation}}
\def\ba{\begin{align}}
\def\ea{\end{align}}

\newcommand{\bem}{\begin{pmatrix}}
\newcommand{\eem}{\end{pmatrix}}

\def\={\;  = \;}
\def\+{\, + \,}

\def\bar{\overline}

\def\rt2{\sqrt{2}}

\title{Spatially Random Disorder in Unitary Fermion System in $(4-\e)$-Dimensions and Effective Action at Finite Temperature}

\author{\small{Rajesh Kumar Gupta,}}
\author{\small{Meenu}}
\affiliation{\small{Department of Physics, Indian Institute of Technology Ropar,
Rupnagar, Punjab 140001, India}}

\emailAdd{rajesh.gupta@iitrpr.ac.in, meenu.20phz0003@iitrpr.ac.in }

\abstract{ Non-relativistic conformal field theory is significant to understand various aspects of an ultra-cold system. In this paper, we study a non-relativistic system of two-component fermions interacting with a complex boson with Yukawa-like interactions near $d=4$-spatial dimensions in the presence of a quenched disorder. The homogeneous theory flows to an interacting fixed point describing a unitary fermion system. In the presence of the disorder, we find that the system has an interesting phase structure in the space of the coupling constants and exhibits an interacting disorder fixed point in $\e$-expansion. The correlation function obeys Lifshitz scaling behaviour at the disorder fixed point with the anisotropic exponent being $z=2+\g_E$. We also study the disorder system at finite temperature and compute the leading contribution to the 1PI effective action.
}

\makeatletter
\gdef\@fpheader{}
\makeatother

\begin{document}

%

\maketitle
\section{Introduction}
Generically, a system of study is pure. A pure system is homogeneous in nature, and the Lagrangian description of such systems involves interactions that enjoy translational invariance. In contrast, most often, a real physical system comes with impurities. These impurities are localized around a point or along a $d$-dimensional surface, breaking the translational invariance of the underlying system.  
Such systems exhibit interesting and sometimes new physical phenomena such as Anderson localization.
These systems are of great interest to condensed matter and statistical physicists, and to study the effect of the impurities on the thermodynamic and critical properties of the system.

Random impurities could arise in many different ways~\cite{Cardy:1996xt}, e.g. a pure system of magnetic ions doped with a diamagnetic atom whose location is random or the Ising model with some random sites left vacant. Depending on the disorder configurations of the underlying system, the impurities could be dynamical or non-dynamical. In other words, while doing statistical mechanics of the impure system, one may or may not need to integrate over the degrees of freedom associated with impurities. Random impurities of dynamical nature are called annealed disorders, while non-dynamical random impurities are called quenched disorders.

As quenched disorder breaks the translational invariance of the system, we can treat these impurities in field theory language by introducing terms in the Lagrangian where the disorder field couples to one or more operators. These couplings are coordinate-dependent and, since these are quenched, destroy the translational symmetry of the original pure system.
One can then compute various thermodynamic quantities like free energy by taking an average over the ensemble of the system with disorder taking values from a probability distribution. It gives quenched average values of the thermodynamic quantities. From the technical perspective, there is also an advantage of averaging. As we will see, it restores the translational invariance of the system, provided the probability distribution for the disorder is also translational invariant. 

The physics of quench disorder and the resultant renormalization group flow have been studied in many different contexts~\cite{Harris_1974, PhysRevB.26.154, DN90a, DN90b, Dotsenko:1994im, Dotsenko:1994sy, Die98, Fujita:2008rs, Hartnoll:2014cua, Hartnoll:2015rza, Aharony:2015aea, Aharony:2018mjm, Narovlansky:2018muj,Gupta:2021plc}. Quenching can be introduced in both classical and quantum systems, which are called classical and quantum disorders, respectively. The field theory analysis of the disorder is often performed using the replica trick, which we will review below. An interesting situation occurs when the system flows to an interacting fixed point with non-zero values of the disorder strength. A characteristic feature of such fixed points in the quantum disorder case is the emergence of Lifshitz scaling, $t\rightarrow \lambda^z t$. Therefore, the physics at the fixed point is generally described by a Lifshitz invariant field theory. Quenched average correlation functions obey the generalized Callan-Symanzik equation, and their scaling behaviour at the fixed point receives anomalous contributions from the dynamical exponent, $z$.
Motivated by these analyses, we would like to investigate the effect of the quantum disorder in the non-relativistic system. We think the investigation might be useful for understanding the critical behaviour of the ultra-cold system in the presence of a disorder.

A system with quench disorder is challenging to solve. The difficulty arises because the averaging is done not at the level of the partition function but at the level of the thermodynamic quantity of interests. This difficulty can be overcome using the replica trick as we now briefly review~\cite{Aharony:2018mjm}.  
Suppose we have a pure system described by an action $S_{0}$. The system is then doped with non-dynamical impurities, which we assume to be described by a disorder field $h(x)$. The disorder field couples to an operator $\mathcal O_{0}(t,x)$. The action of the disordered system is
\be
S=S_{0}+\int dt\,d^{d}x\,h(x)\mathcal O_{0}(t,x)\,.
\ee
Note that the quantum disorder, $h(x)$, is homogeneous in time. We then consider an ensemble of the disordered system where the probability of finding a specific profile of the disorder field is given by $P[h]$.
The spatial randomness implies that the distribution $P[h]$ is such that the spatial correlation of the disorder field is much smaller than the intrinsic length scale of the system, e.g. lattice spacing. A convenient choice to achieve this is the Gaussian distribution,
\be\label{Gaussiandistr}
P[h]=\mathcal N\,e^{-\frac{1}{2v}\int d^{d}x\,h^{2}(x)}\,.
\ee
Now, the partition function (or the generating function when we include sources) for a given profile of the disorder field is
\be
Z[h]=e^{W[h]}=\int [d\eta^{\dagger}][d\eta]\,e^{iS_{0}(\eta,\eta^{\dagger})+i\int dt\,d^{d}x\,h(x)\mathcal O_{0}(t,x)}\,.
\ee
The above is the partition function of a system in the ensemble for a specific disorder profile. Connected correlation functions for a given disorder profile are obtained by computing the derivatives of $W[h]$ (in the presence of sources). The quenched average correlation function is obtained by averaging over the disorder distribution. For example, the quenched average free energy (and its derivatives) is given by,
\be
\bar F=\int [dh]P[h]W[h]\,.
\ee
Since the distribution~\eqref{Gaussiandistr} is translational invariant, the quenched average quantities are also translational invariant.

The quenched average free energy (and its derivatives) can be calculated using the replica trick. In the replica trick, we introduce $\underline{n}$-copies of the original theory with the same profile of the disorder field. The fields and the disorder operator are now labelled by replica index $A$, where $A=1,...\underline{n}$. The partition function is then defined as,
\be
Z^{\underline{n}}[h]=\int \prod_{A=1}^{\underline{n}}[d\eta_{A}^{\dagger}][d\eta_{A}]e^{i\sum_{A=1}^{\underline{n}}S_{0,A}(\eta_{A},\eta_{A}^{\dagger})+i\int dt\,d^{d}x\,h(x)\sum_{A=1}^{\underline{n}}\mathcal O_{0,A}(t,x)}\,.
\ee
The above partition function is computed for the integer $\underline{n}$. Assuming that the analytic continuation exists for real values of $\underline{n}$, the quenched average free energy is given by
\bea
\bar F&=&\lim_{\underline{n}\rightarrow 0}\frac{\p}{\p \underline{n}}\int D[h]P[h]Z^{\underline{n}}[h]\,,\nn\\
&=&\lim_{\underline{n}\rightarrow 0}\frac{\p}{\p \underline{n}}\int \prod_{A=1}^{\underline{n}}[d\eta_{A}^{\dagger}][d\eta_{A}]\,e^{i\sum_{A=1}^{\underline{n}}S_{0,A}(\eta_{A},\eta_{A}^{\dagger})-\frac{v}{2}\int dt\,dt'\,d^{d}x\,\sum_{A,B=1}^{\underline{n}}\mathcal O_{0,A}(t,x)\mathcal O_{0,B}(t',x)}\,,\nn\\
&=&\lim_{\underline{n}\rightarrow 0}\frac{\p}{\p \underline{n}}\int \prod_{A=1}^{\underline{n}}[d\eta_{A}^{\dagger}][d\eta_{A}]\,e^{iS_{\text{repl.}}}\,,
\eea
where the action for the replicated theory is 
\be\label{ReplicatedAction}
S_{\text{repl.}}=\sum_{A=1}^{\underline{n}}S_{0,A}(\eta_{A},\eta_{A}^{\dagger})+\frac{iv}{2}\int dt\,dt'\,d^{d}x\,\sum_{A,B=1}^{\underline{n}}\mathcal O_{0,A}(t,x)\mathcal O_{0,B}(t',x)\,.
\ee
Thus, we see that the replica trick reduces the computation of the quenched average free energy (and its derivatives) to a path integral computation, but with a price that the fields with different replica indexes couple with each other. We also observe a couple of the effects of averaging over the probability distribution: Firstly, the replicated action is complex, and secondly, it is non-local in time. 

We will be interested in the quenched average correlation functions and the Callan-Symanzik equation satisfied by these. The quenched average correlation function is given by
\be
\bar{<\mathcal O(t_{1},x_{1}).....\mathcal O(t_{\ell},x_{\ell})>}=\int [dh]P[h]<\mathcal O(t_{1},x_{1}).....\mathcal O(t_{\ell},x_{\ell})>_{h}\,,
\ee
where the correlation function $<\mathcal O(t_{1},x_{1}).....\mathcal O(t_{\ell},x_{\ell})>_{h}$ is computed for a given profile of the disorder field. Following the replica trick, one can express the quenched average connected correlation function in terms of the correlation function computed in the replicated theory. More precisely, the relation is 
\be\label{AverageCorr.}
\bar{<\mathcal O(t_{1},x_{1}).....\mathcal O(t_{\ell},x_{\ell})>_{\text{conn.}}}=\lim_{\underline{n}\rightarrow 0}\frac{\p}{\p \underline{n}}<\mathcal \sum_{A_{1}}O_{A_{1}}(t_{1},x_{1}).....\sum_{A_{\ell}}\mathcal O_{A_{\ell}}(t_{\ell},x_{\ell})>_{\text{repl.}}\,.
\ee
The above relation makes it simpler to compute the average correlation function. It amounts to computing the correlation function in the replicated theory using the standard tools of the perturbative quantum field theory, and then analytically continuing the result for real values of $\underline{n}$.
Note that the RHS of~\eqref{AverageCorr.} need not be the connected correlation function. 

The quenched average connected correlation function obeys generalized Callan Symanzik equation. This can be obtained starting from the relation~\eqref{AverageCorr.}.
We begin with the Callan-Symanzik equation satisfied by the renormalized Green's function in the replicated theory (we assume that the operators $O_{A}(t,x)$ do not mix with any other operator)
\be
\Big(\m\frac{\p}{\p\m}+\beta_{i}\frac{\p}{\p g_{i}}+\ell\,\g_{\mathcal O}\Big)<\mathcal O_{A_{1}}(t_{1},x_{1}).....\mathcal O_{A_{\ell}}(t_{\ell},x_{\ell})>^{\text{renor.}}_{\text{repl.}}=0\,,
\ee
where $\m$ is the renomalization scale, $\beta_{i}$ are beta functions for all the coupling constants and $\g_{\mathcal O}$ is the anomalous dimension. Then we have
\bea
&&\Big(\m\frac{\p}{\p\m}+\beta_{i}\frac{\p}{\p g_{i}}+\ell\,\g_{\mathcal O}\Big)\frac{\p}{\p \underline {n}}<\sum_{A_{1}}\mathcal O_{A_{1}}(t_{1},x_{1}).....\sum_{A_{\ell}}\mathcal O_{A_{\ell}}(t_{\ell},x_{\ell})>^{\text{renor.}}_{\text{repl.}}\nn\\
&&+\Big(\m\frac{\p}{\p\m}+\frac{\p\beta_{i}}{\p \underline {n}}\frac{\p}{\p g_{i}}+\ell\,\frac{\p\g_{\mathcal O}}{\p \underline {n}}\Big)<\sum_{A_{1}}\mathcal O_{A_{1}}(t_{1},x_{1}).....\sum_{A_{\ell}}\mathcal O_{A_{\ell}}(t_{\ell},x_{\ell})>^{\text{renor.}}_{\text{repl.}}=0\,.
\eea
Since the replica symmetry implies that $<\sum_{A_{1}}\mathcal O_{A_{1}}(t_{1},x_{1}).....\sum_{A_{\ell}}\mathcal O_{A_{\ell}}(t_{\ell},x_{\ell})>^{\text{renor.}}_{\text{repl.}}$ will be at least proportional to $\underline {n}$, we see that when we take the limit $\lim \underline {n}\rightarrow 0$, we obtain
\be
\Big(\m\frac{\p}{\p\m}+\lim_{\underline {n}\rightarrow 0}\beta_{i}\frac{\p}{\p g_{i}}+\ell\,\lim_{\underline {n}\rightarrow 0}\g_{\mathcal O}\Big)\bar{<\mathcal O(t_{1},x_{1}).....\mathcal O(t_{\ell},x_{\ell})>_{\text{conn.}}^{\text{renor.}}}=0
\ee
Thus, the beta function and the anomalous dimension of the disordered theory are $\lim\limits_{\underline {n}\rightarrow 0}\beta_{i}$ and $\lim\limits_{\underline {n}\rightarrow 0}\g_{\mathcal O}$. In particular, the fixed point of the resultant beta function gives the disordered fixed point. 

In the present paper, our goal would be to study the effects of the disorder on the physical properties of non-relativistic systems~\footnote{Spatially random disorder and its effect on the transport properties of the system has been studied previously, e.g. see~\cite{Patel:2022gdh} and references therein.}. In this direction, we study a system of two-component fermions and a complex boson interacting with Yukawa-like interactions near $d=4$-spatial dimensions in the presence of a quenched disorder. The original homogeneous system has been studied in the past in the context of the unitary fermi system~\cite{PhysRevA.75.033608, Nishida:2007pj}, and exhibits an interacting IR fixed point in $\e$-expansion that is described by a non-relativistic conformal field theory~\footnote{The subject of non-relativistic conformal field theory is of great interest in high energy, condensed matter and statistical physics. For various aspects of non-relativistic conformal field theory, see~\cite{Hagen:1972pd, Henkel:1993sg, Mehen:1999nd, Nishida:2006br, Nishida:2006eu, Nishida:2007pj, Braaten:2008uh, Golkar:2014mwa, Goldberger:2014hca, Gupta:2022azd}. }. We also study the finite temperature aspects of the disorder system. We compute the leading contribution to the free energy in $(4-\e)$-dimensions, which requires summing all the 1PI diagrams. The free energy can then be used to compute the thermodynamic quantities of the disorder system.

The organization of the paper is as follows: In section~\ref{NRCFT}, we review the renormalization group flow in a non-relativistic field theory of two-component fermions interacting with a complex boson in $(4-\e)$-spatial dimensions. The flow has an interacting fixed point which describes fermions at unitarity. In section~\ref{DisorderDeformation}, we consider the deformation of the non-relativistic system by a spatially random disorder. We study the effect of the disorder in the renormalization group flow of the homogeneous system. We find that the system has an interesting phase structure in the space of the coupling constants. The correlation function obeys Lifshitz scaling behaviour at the disorder fixed point with the anisotropic exponent being $z=2+\g_E$. In section~\ref{Finite Temp. phases}, we investigate the disorder system at a finite temperature. We compute the leading contribution to the free energy in $(4-\e)$-spatial dimensions, which is relevant for the thermodynamic study of the system. The free energy computation is done in two different limits: in the first case, we consider the coupling constants independent of $\e$, and in the second case, we consider the coupling constant $\mathcal O(\e)$. In both cases, we can compute the leading contribution to the free energy.
\section{A non-relativistic conformal field theory}\label{NRCFT}
This section considers a non-relativistic field theory that flows to an interacting fixed point. It is a theory of a 2-component fermion and a complex scalar field interacting through a Yukawa interaction in $(4-\e)$-spatial dimensions. At the fixed point, we have a non-relativistic conformal field theory.

\begin{figure}[htpb]
\begin{center}
\vspace{-8cm}
\centering
\includegraphics[width=6in]{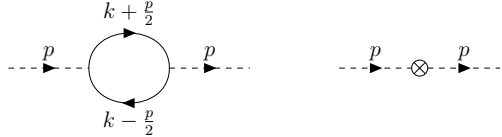}
\vspace{-9 cm}
\caption{Scalar two point function with the wave function renormalization counter term.}\label{Scalar2-pointFn1}
\end{center}
\end{figure}

Let us start with a theory of 2-component fermions and a complex scalar in $d$-spatial dimensions~\cite{Nishida:2007pj, PhysRevA.75.033608}\footnote{Such a system has also been studied in the presence of the boundary, see\cite{Gupta:2022azd}.}. The action is 
\be\label{NRFT}
S=\int dt\,d^{d}x\Big[\psi_{\sigma}^{\dagger}(i\p_{t}+\frac{\nabla^{2}}{2})\psi_{\sigma}\Big]+\int dt\,d^{d}x\Big[\phi^{\dagger}(i\p_{t}+\frac{\nabla^{2}}{4})\phi\Big]+\int dt\,d^{d}x\Big[g_{0}\psi_{\uparrow}^{\dagger}\psi_{\downarrow}^{\dagger}\phi+g_{0}\psi_{\downarrow}\psi_{\uparrow}\phi^{\dagger}\Big]\,.
\ee
In terms of the two component fermion, 
\be
\Psi=\begin{pmatrix}\psi_{\uparrow}\\\psi_{\downarrow}^{\dagger}\end{pmatrix},\quad \Psi^{\dagger}=\begin{pmatrix}\psi^{\dagger}_{\uparrow}&\psi_{\downarrow}\end{pmatrix}\,,
\ee
the above action is
\be
S=\int dt\,d^{d}x\Big[\Psi^{\dagger}(i\p_{t}+\sigma_{3}\frac{\nabla^{2}}{2})\Psi+\phi^{\dagger}(i\p_{t}+\frac{\nabla^{2}}{4})\phi+g_{0}\Psi^{\dagger}\sigma_{+}\Psi \phi+g_{0}\Psi^{\dagger}\sigma_{-}\Psi\phi^{\dagger}\Big]\,,
\ee
where the matrices are given by
\be
\sigma_{+}=\frac{1}{2}(\sigma_{1}+i\sigma_{2})=\begin{pmatrix}0&1\\0&0\end{pmatrix},\quad \sigma_{-}=\frac{1}{2}(\sigma_{1}-i\sigma_{2})=\begin{pmatrix}0&0\\1&0\end{pmatrix}\,.
\ee
The coupling constant $g_{0}$ is marginal in $d=4$ and relevant in $d=(4-\e)$-spatial dimensions. We will compute the $\beta$-function at one loop in $\e$-expansion.
For the loop computations, we need the following propagator:
\bea
&&\Big<\mathcal T\Psi(t,x)\Psi^{\dagger}(t',x')\Big>=\int \frac{dE\,d^{d}p}{(2\pi)^{d+1}}e^{-iE(t-t')-i\vec p\cdot(\vec x-\vec x')}G(E,p),\nn\\
&&\Big<\mathcal T\phi(t,x)\phi^{\dagger}(t',x')\Big>=\int \frac{dE\,d^{d}p}{(2\pi)^{d+1}}e^{-iE(t-t')-i\vec p\cdot(\vec x-\vec x')}D(E,p)\,.
\eea
Here
\bea
G(E,p)=\frac{i}{E^{2}-\epsilon_{\vec p}^{2}+i\delta^{+}}\begin{pmatrix}E+\epsilon_{\vec p}&0\\0&E-\epsilon_{\vec p}\end{pmatrix},\quad D(E,p)=\frac{i}{E-\frac{\epsilon_{\vec p}}{2}+i\delta^{+}}
\eea
and
\be
\epsilon_{\vec p}=\frac{\vec p^{2}}{2}\,.
\ee
We find that the theory exhibits IR fixed point in $\e$-expansion. We see this as follows: There is no self-energy correction to fermion. The 1PI correction to the three-point function vanishes at one loop. Thus, the coupling constant $g_{0}$ is renormalized only through the renormalization of the scalar kinetic term. Replacing $\phi\rightarrow \sqrt{Z_{\phi}}\,\phi$ and defining $g=g_{0}\sqrt{Z_{\phi}}$, the one loop self energy contribution to the scalar kinetic term is (see the figure~\ref{Scalar2-pointFn1})
\bea
&&g^{2}\int\frac{dE\,d^{d}k}{(2\pi)^{d+1}}G_{11}(E+\frac{p_{0}}{2},\vec k+\frac{\vec p}{2})G_{22}(E-\frac{p_{0}}{2},\vec k-\frac{\vec p}{2})+i\delta_{\phi}(p_{0}-\frac{\vec p^{2}}{4})\nn\\
&&\qquad\qquad\quad\quad=\,-ig^{2}\int\frac{d^{d}k}{(2\pi)^{d}}\frac{1}{p_{0}-\frac{\vec p^{2}}{4}-\vec k^{2}}+i\delta_{\phi}(p_{0}-\frac{\vec p^{2}}{4})\nn\\
&&\qquad\qquad\qquad=\,-ig^{2}\frac{\pi^{\frac{d}{2}}}{\Gamma(\frac{d}{2})(2\pi)^{d}}(-1)^{\frac{d}{2}}\Big(p_{0}-\frac{\vec p^{2}}{4}\Big)^{\frac{d}{2}-1}\frac{\pi}{\sin\frac{\pi d}{2}}+i\delta_{\phi}(p_{0}-\frac{\vec p^{2}}{4})\,.
\eea
In the above $Z_{\phi}=1+\delta_{\phi}$. Differentiating the above with respect to $p_{0}$ and requiring the divergence to cancel, we obtain
\be\label{DeltaPhiscalar}
\delta_{\phi}=-\frac{g^{2}}{8\pi^{2}\e}\,.
\ee
Next, we solve the Callan-Symanzik equation for two and three point functions. Solving these, we obtain 
\be
\beta=-\frac{\e}{2}g+\frac{g^{3}}{16\pi^{2}}\,,\quad \text{and}\quad\g_{\phi}=\frac{g^{2}}{16\pi^{2}}\,.
\ee
We see from the expression of the beta function that, other than the Gaussian fixed point, there is a non-trivial interacting fixed point at
\be
g_{*}=2\pi\sqrt{2\e}\,.
\ee
At the fixed point, we have a non-relativistic conformal field theory that describes the system of unitary fermions.

The above computation can also be extended to an arbitrary number of fermions. 
The action for the $N$-number of fermions interacting with a complex scalar field is, 
\be
S=\int dt\,d^{d}x\Big[\Psi_{i}^{\dagger}(i\p_{t}+\sigma_{3}\frac{\nabla^{2}}{2})\Psi_{i}\Big]+\phi^{\dagger}(i\p_{t}+\frac{\nabla^{2}}{4})\phi+g_{0}\Psi_{i}^{\dagger}\sigma_{+}\Psi_{i} \phi+g_{0}\Psi_{i}^{\dagger}\sigma_{-}\Psi_{i}\phi^{\dagger}\Big]\,,
\ee
where the repeated index is summed over. In this case, we get
\be
\delta_{\phi}=-\frac{N}{8\pi^{2}\epsilon}g^{2}\,,
\ee
and the $\beta$ function is
\be
\beta_{g}=-\frac{\epsilon g}{2}+\frac{Ng^{3}}{16\pi^{2}}\,.
\ee
The fixed point is at $g_{*}=2\pi\sqrt{\frac{2\e}{N}}$.

At the fixed point, we have a non-relativistic conformal field theory, and we can compute the scaling dimension of various operators, labelled by the particle number, e.g. the scaling dimension of the one and two-fermion operators are
\be
\Delta_{\psi}=\frac{d}{2}=2-\frac{\e}{2},\quad \Delta_{\phi}=\frac{d}{2}+\g_{\phi}=2\,.
\ee
The scaling dimensions would be the Hamiltonian's energy eigenvalue for the system of fermions in a harmonic trap.
\section{Spatially random deformation of fermions at unitarity}\label{DisorderDeformation}
Next, we deform the above action~\eqref{NRFT} by a spatially random disorder. The simplest but non-trivial situation is where the disorder field couples to a scalar operator, which respects the particle number symmetry. These operators are constructed out of the basic fields of the homogeneous theory. For the unitary fermion system discussed above, these operators are $\phi^{\dagger}\phi, \Psi^{\dagger}\Psi$ and $\Psi^{\dagger}\sigma_{3}\Psi$. 
From the dimension analysis, we find that the disorder strength, $v$, has the dimension
\be
[v]=4-d\,.
\ee
Thus, the disorder deformation is marginal in $d=4$ and relevant in $d=(4-\e)$-dimensions. We will work in the dimension $d=4-\e$. These deformations will trigger a renormalization group flow in the space of all coupling constants, including the disorder strength $v$, leading to a new fixed point. 
We will consider each of these deformations separately in the discussion below and study the corresponding renormalization group flow.
\subsection{Deformation by the disorder operator $\phi^{\dagger}\phi$}\label{DisorderOperatorPhi}
We will begin with the disorder operator $\phi^{\dagger}\phi$.
The deformed action is
\bea
S&=&\int dt\,d^{d}x\Big[\Psi^{\dagger}(i\p_{t}+\sigma_{3}\frac{\nabla^{2}}{2})\Psi\Big]+\int dt\,d^{d}x\Big[\phi^{\dagger}(i\p_{t}+\frac{\nabla^{2}}{4})\phi\Big]\nn\\
&&+\int dt\,d^{d}x\Big[g_{0}\Psi^{\dagger}\sigma_{+}\Psi \phi+g_{0}\Psi^{\dagger}\sigma_{-}\Psi\phi^{\dagger}\Big]
+\int dt\,d^{d}x\,h(x)\phi^{\dagger}\phi\,.
\eea

After integrating over the disorder field, we obtain the replicated action to be
\bea
S^{(b)}_{\text{repl.}}=\sum_{A=1}^{\underline {n}}S_{0,A}+i\frac{v}{2}\int d^{d}x\,dt\,dt'\,\sum_{A,B=1}^{\underline {n}}\phi_{A}^{\dagger}(x,t)\phi_{A}(x,t)\phi_{B}^{\dagger}(x,t')\phi_{B}(x,t')\,,
\eea
where
\bea\label{Replic.PureAction}
S_{0,A}&=&\int dt\,d^{d}x\Big[\Psi_{A}^{\dagger}(i\p_{t}+\sigma_{3}\frac{\nabla^{2}}{2m})\Psi_{A}\Big]+\int dt\,d^{d}x\Big[\phi_{A}^{\dagger}(i\p_{t}+\frac{\nabla^{2}}{4m})\phi_{A}\Big]\nn\\
&&+\int dt\,d^{d}x\Big[g_{0}\Psi_{A}^{\dagger}\sigma_{+}\Psi_{A} \phi_{A}+g_{0}\Psi_{A}^{\dagger}\sigma_{-}\Psi_{A}\phi_{A}^{\dagger}\Big]\,.
\eea
We see from the above action that the replicated theory is not local in time.

Next, we would like to find the Callan-Symanzik equation satisfied by the quenched average renormalized correlation functions. Following the standard method of renormalization, we introduce the counter terms by replacing the fields $\phi_{A}\rightarrow \sqrt{Z_{\phi}}\phi_{A}$ and $\Psi_{A}\rightarrow \sqrt{Z_{\Psi}}\Psi_{A}$. The replicated action becomes
\bea
S^{(b)}_{\text{repl.}}=\sum_{A=1}^{\underline {n}}S_{0,A}+i\frac{\m^{\e}\lambda(1+\delta_{\lambda})}{2}\int d^{d}x\,dt\,dt'\,\sum_{A,B=1}^{\underline {n}}\phi_{A}^{\dagger}(x,t)\phi_{A}(x,t)\phi_{B}^{\dagger}(x,t')\phi_{B}(x,t')\,,\nn\\
\eea
where
\bea
S_{0,A}&=&\int dt\,d^{d}x\Big[Z_{\psi}\Psi_{A}^{\dagger}(i\p_{t}+\sigma_{3}\frac{\nabla^{2}}{2m})\Psi_{A}\Big]+\int dt\,d^{d}x\Big[Z_{\phi}\phi_{A}^{\dagger}(i\alpha\p_{t}+\frac{\nabla^{2}}{4m})\phi_{A}\Big]\nn\\
&&+\int dt\,d^{d}x\,g\m^{\frac{\e}{2}}(1+\delta_{g})\Big[\Psi_{A}^{\dagger}\sigma_{+}\Psi_{A} \phi_{A}+\Psi_{A}^{\dagger}\sigma_{-}\Psi_{A}\phi_{A}^{\dagger}\Big]\,.
\eea
In the above we have introduced the renormalized coupling constants as $Z_{\phi}^{2}v=\m^{\e}\lambda(1+\delta_{\lambda})$ and $Z_{\Psi}\sqrt{Z_{\phi}}g_{0}=\m^{\frac{\e}{2}}g(1+\delta_{g})$. As we will see below, the loop corrections also generate a term proportional to $\phi^{\dagger}\p_{t}\phi$. Therefore, there is a renormalization of the frequency part of the kinetic term of the scalar action. To take it into account, we introduce the counters term $Z_{\phi}\alpha\,\phi^{\dagger}\p_{t}\phi=(1+\delta\alpha+\delta_{\phi})\phi^{\dagger}\p_{t}\phi$.

\begin{figure}[htpb]
\begin{center}
\vspace{-8cm}
\centering
\includegraphics[width=6in]{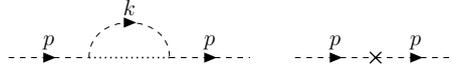}
\vspace{-8 cm}
\caption{Scalar two point function with the energy counter term. The dotted lines represent the disorder vertex. }\label{Scalar2ptwithDisorder}
\end{center}
\end{figure}

Next, we want to compute the $\beta$-function at one loop order. We need the following propagator:
\bea
&&\Big<\mathcal T\Psi_{A}(t,x)\Psi_{B}^{\dagger}(t',x')\Big>=\delta_{A,B}\int \frac{dE\,d^{d}p}{(2\pi)^{d+1}}e^{-iE(t-t')-i\vec p\cdot(\vec x-\vec x')}G(E,p),\nn\\
&&\Big<\mathcal T\phi_{A}(t,x)\phi_{B}^{\dagger}(t',x')\Big>=\delta_{A,B}\int \frac{dE\,d^{d}p}{(2\pi)^{d+1}}e^{-iE(t-t')-i\vec p\cdot(\vec x-\vec x')}D(E,p)\,.
\eea
As in the homogeneous case, there is no contribution to the fermion's two-point function at one loop; thus, there is no wave function renormalization for fermions, i.e. $Z_{\psi}=1$. 
Next, we compute the two-point function for scalar, i.e. $<T\phi_{A}(t,x)\phi^{\dagger}_{B}(t',x')>$ in the replicated theory. At one loop, we have two contributions. The first contribution is due to the fermions in the loop as in the section~\ref{NRCFT}, which determines $Z_{\phi}$. The second contribution comes from the disorder vertex~\footnote{Note that one more contribution comes from the self-contractions of the disorder vertex. The contribution is $\sim \underline{n}\,\delta_{A,B}\lambda \int dt\,\int \frac{de\,d^{d}q}{(2\pi)^{d+1}}D(e,\vec q)$. The contribution is proportional to $\underline{n}$ and, therefore, does not contribute to the disorder average two-point function. We also note that there is IR divergence due to the volume of the time.}, shown in the figure~\ref{Scalar2ptwithDisorder}~\footnote{All the Feynman diagrams in the paper have been drawn using Tikz-Feynman~\cite{Ellis:2016jkw}.}, and is given by
\bea
&&-\lambda\,\delta_{A,B}\int\frac{d^{d}q}{(2\pi)^{d}}\frac{i}{E_{1}-\frac{q^{2}}{4}+i\delta}+i\delta_{\alpha}\,E_{1}\delta_{A,B}\nn\\
&&\qquad\qquad\qquad=\,-\lambda\,\delta_{A,B}\frac{2\pi^{\frac{d}{2}}}{\Gamma(\frac{d}{2})(2\pi)^{d}}\Big(2i(-1)^{\frac{4-d}{2}}(4E_{1})^{\frac{d-2}{2}}\frac{\pi}{\sin\frac{\pi d}{2}}\Big)+i\delta_{\alpha}\,E_{1}\delta_{A,B}\,.
\eea
In the above $E_{1}$ is the energy associated with the external lines.
Cancellation of the divergence requires that
\be\label{DeltaAlphaScalar}
\delta_{\alpha}=-\frac{2\lambda}{\pi^{2}\e}\,.
\ee

\begin{figure}[htpb]
\begin{center}
\vspace{-4cm}
\centering
\includegraphics[width=6in]{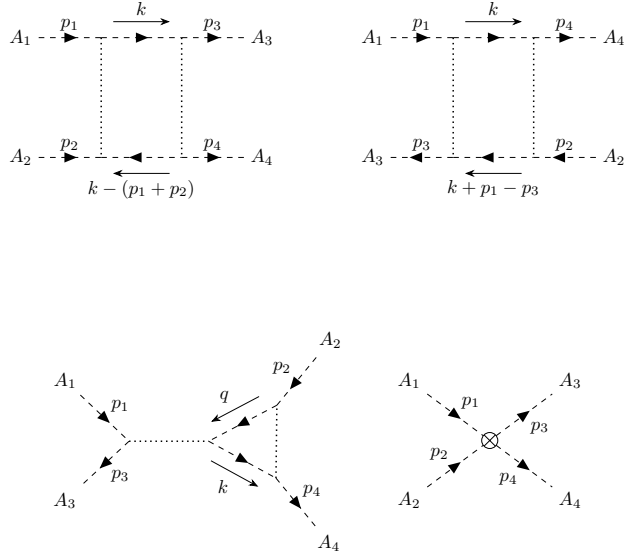}
\vspace{-8 cm}
\caption{Scalar four point function with the counter term. The dotted lines represent the disorder vertex.}\label{Scalar4ptDisorder}
\end{center}
\end{figure}

\begin{figure}[htpb]
\begin{center}
\vspace{-0.3 cm}
\includegraphics[width=4in]{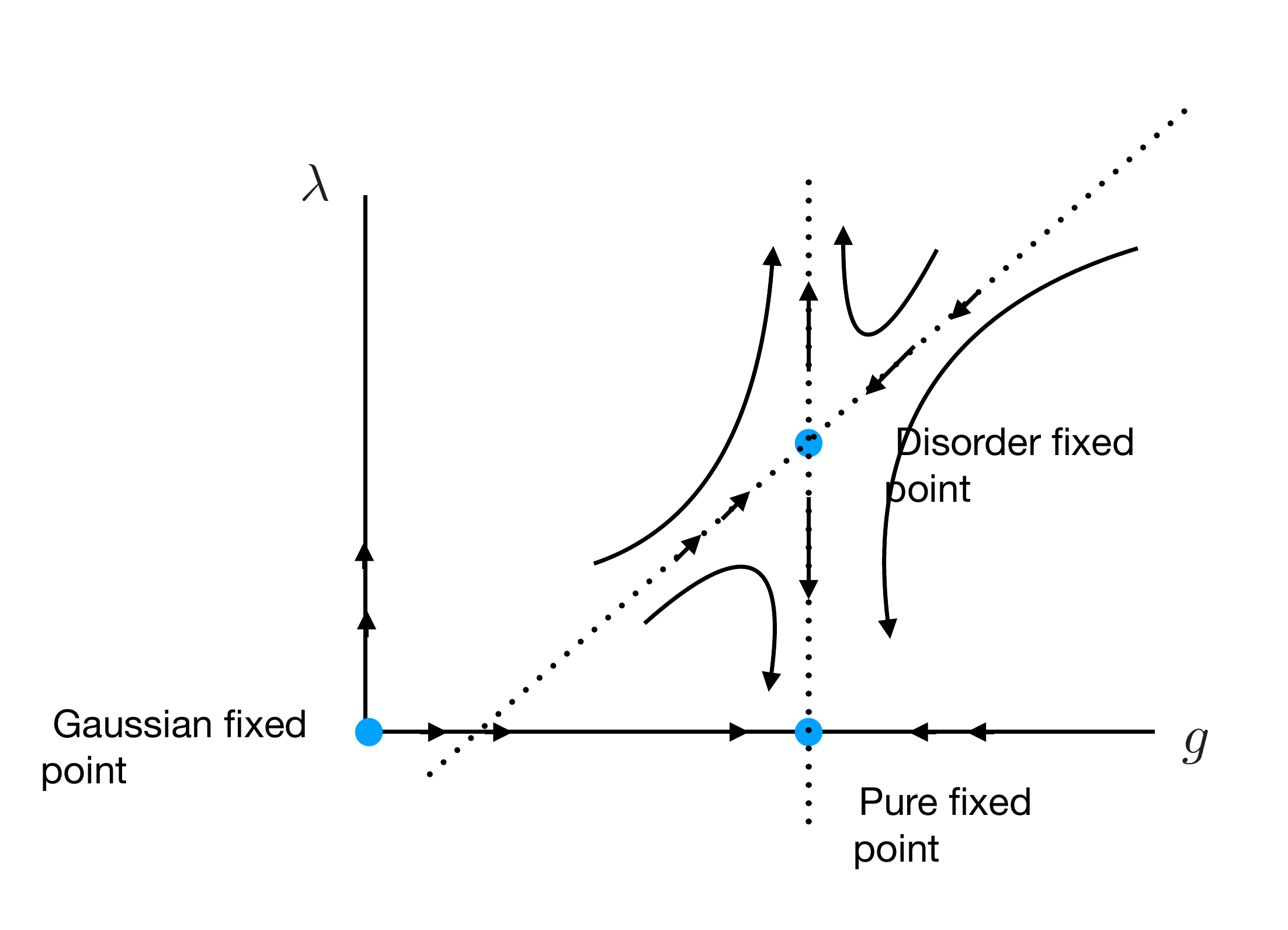}
\vspace{-0.5cm}
\caption{Renormalization group flow in the space of coupling constants.\label{PhaseDiagram}}
\end{center}
\end{figure}

Next, we compute the four-point function $<T\phi^{\dagger}_{A_{1}}(t_{1},x_{1})\phi^{\dagger}_{A_{2}}(t_{2},x_{2})\phi_{A_{3}}(t_{3},x_{3})\phi_{A_{4}}(t_{4},x_{4})>$ at one loop order. There are two contributions. The first contribution is due to the scalar in the loop as shown in the figure~\ref{Scalar4ptDisorder}. 
The contribution is given by~\footnote{There is another channel that give rise to the contribution proportional to $\delta_{A_{1},A_{4}}\delta_{A_{2},A_{3}}$, however, for the $\beta$-function calculation an explicit expression for one is enough.}
\bea
\delta_{A_{1},A_{3}}\delta_{A_{2},A_{4}}&&\Big[-\lambda\,\delta\lambda+\lambda^{2}\int \frac{d^{d}q}{(2\pi)^{d}}\Big(D(e_{1},\vec q)D(e_{2},\vec p_{1}+\vec p_{2}-\vec q)\nn\\
&&+D(e_{4},\vec q)D(e_{2},\vec p_{1}-\vec p_{3}+\vec q)
+D(e_{3},\vec q)D(e_{3},\vec p_{1}-\vec p_{3}+\vec q)\nn\\
&&+D(e_{1},\vec q)D(e_{4},\vec p_{1}-\vec p_{4}+\vec q)\Big)\Big]\,.
\eea
In the above $(e_{i},\vec p_{i})$ for $i=1,..4$ are the energy and momentum associated with the external legs. We have not shown the delta functions that enforce energy and momentum conservation. Note that there is one more contribution which is proportional to $\underline {n}\times\delta (E=0)$ and is volume (temporal) divergent. Since we are interested in computing the beta function, which is sensitive to UV divergence, we will ignore the volume-divergent contribution. \\
The second contributions come from the diagram with fermions running in the loop. The contribution is given by 
\bea
&&\frac{g^{4}}{4!}\delta_{A_{1},A_{3}}\delta_{A_{2},A_{4}}\int \frac{dE\,d^{d}q}{(2\pi)^{d+1}}G_{11}(E,\vec k)G_{11}(E+e_{2}-e_{4},\vec k+\vec p_{2}-\vec p_{4})\nn\\
&&\qquad\qquad\qquad\qquad\times G_{22}(E-e_{1},\vec k-\vec p_{1})G_{22}(E-e_{4},\vec k-\vec p_{4})\,.
\eea
Evaluating the above, we find that the integral is finite in $d=4$. Finally, we obtain the counter term by 
requiring that the renormalized four-point function is finite in $d=4-\e$, and is given by~\footnote{In the paper, we are working in minimal subtraction scheme.} 
\be\label{deltaLambScalar}
\delta_{\lambda}=-\frac{8\lambda}{\pi^{2}\epsilon}\,.
\ee
Thus, we get the $\beta$ function to be
\bea\label{BetaFnScalar}
&&\beta_{g}=-\frac{\epsilon g}{2}+\frac{g^{3}}{16\pi^{2}}\,,\nn\\
&&\beta_{\lambda}=-\epsilon\lambda-\frac{8\lambda^{2}}{\pi^{2}}+\frac{\lambda g^{2}}{4\pi^{2}}\,.
\eea
The physical fixed points are obtained by solving $\beta_{g}=0=\beta_{\lambda}$ and are as follows:
\bea
(\lambda_{*}, g_{*})=(0,0),(0,2\pi\sqrt{2\epsilon{}}), (\frac{\pi^{2}\epsilon}{8},2\pi\sqrt{2\epsilon})\,.
\eea
Note that there is a non trivial interacting IR fixed point which is a disorder fixed point and given by
\be
(\lambda_{*}, g_{*})=(\frac{\pi^{2}\epsilon}{8},2\pi\sqrt{2\epsilon}).
\ee
We can analyse the stability of these fixed points in the linearized approximation. Defining $g=g_{*}+\delta g$ and $\lambda=\lambda_{*}+\delta\lambda$, we get
\be
\m\frac{d}{d\m}\begin{pmatrix}\delta g\\\delta\lambda\end{pmatrix}=\begin{pmatrix}\frac{1}{16}\Big(\frac{3g_{*}^{2}}{\pi^{2}}-8\e\Big)&&0\\\frac{g_{*}\lambda_{*}}{2\pi^{2}}&&\frac{g_{*}^{2}-4\pi^{2}\e-64\lambda_{*}}{4\pi^{2}}\end{pmatrix}\begin{pmatrix}\delta g\\\delta\lambda\end{pmatrix}\,.
\ee
We now make the following observations about the stability of the various fixed points: At the Gaussian fixed point, the eigenvalues of the above matrix are $(-\frac{\e}{2},-\e)$. Thus, for $d<4$, the Gaussian fixed point is an unstable fixed point, and both deformations are relevant.
At the pure fixed point, the eigenvalues are $(\e,\e)$. Thus, both the deformation away from the pure fixed points are irrelevant deformation. Therefore, the pure fixed point is an attractive fixed point. The disorder fixed point has an interesting stability structure. The eigenvalues of the matrix are $(-\e,\e)$. The eigenvalue $+\e$ is associated with the direction $g=2\pi\sqrt{2\epsilon}$ while the eigenvalue $-\e$ is associated with the direction $\lambda+\frac{\pi^{2}\e}{8}=\frac{\pi\sqrt{\e}}{8\sqrt{2}}g$. The deformation in the direction $g=2\pi\sqrt{2\epsilon}$ is the relevant one while the deformation in the direction $\lambda+\frac{\pi^{2}\e}{8}=\frac{\pi\sqrt{\e}}{8\sqrt{2}}g$ is the irrelevant. These results are plotted in figure\,\ref{PhaseDiagram}. Thus, we see three distinct situations here. A theory starting from the point below the line $\lambda+\frac{\pi^{2}\e}{8}=\frac{\pi\sqrt{\e}}{8\sqrt{2}}g$ will flow to the pure fixed point describing the unitary fermi system, whereas starting from above the line will make the system completely disordered. However, the points on the line will flow to the Wilson-Fisher fixed point.

We can generalize the above by including $N$-number of fermions. In this case we have the $\beta$ functions given by
\bea
&&\beta_{g}=-\frac{\epsilon g}{2}+\frac{Ng^{3}}{16\pi^{2}}\,,\nn\\
&&\beta_{\lambda}=-\epsilon\lambda-\frac{8\lambda^{2}}{\pi^{2}}+\frac{N\lambda g^{2}}{4\pi^{2}}\,.
\eea
The physical fixed points are as follows:
\bea\label{CriticalPt}
(\lambda_{*}, g_{*})=(0,0),(0, \frac{2\pi\sqrt{2\epsilon}}{\sqrt{N}}), (\frac{\pi^{2}\epsilon}{8}, \frac{2\pi\sqrt{2\epsilon}}{\sqrt{N}})\,.
\eea
Thus, we see that other than the Gaussian and pure fixed point, there is a non-trivial disorder fixed point at
\be
(\lambda_{*}, g_{*})=\Big(\frac{\pi^{2}\epsilon}{8},\frac{2\pi\sqrt{2\epsilon}}{\sqrt{N}}\Big)\,.
\ee

Finally, we would like to comment on the Callan-Symanzik equation satisfied by the quenched average renormalized correlation function. We find that the 2-point function satisfies
\be\label{DisCallanSymanzik}
\Big(\m\frac{\p}{\p\m}+\beta_{g}\frac{\p}{\p g}+\beta_{\lambda}\frac{\p}{\p\lambda}-\g_{E}\,E\frac{\p}{\p E}+2\g_{\phi}\Big)\tilde D(E,\vec p)=0\,,
\ee
where
\be
\overline{<\phi(E_{1},\vec p_{1})\phi^{\dagger}(E_{2},\vec p_{2})>_{\text{conn.}}}=(2\pi)^{d+1}\delta(E_{1}-E_{2})\delta^{d}(\vec p_{1}-\vec p_{2})\tilde D(E_{1},\vec p_{1})\,,
\ee
and
\be
\g_{E}=\frac{2\lambda}{\pi^{2}}\,.
\ee
If we restrict to the fixed point in the space of the coupling constant, then the Callan-Symanzik equation~\eqref{DisCallanSymanzik} in terms of space and time coordinate is,
\be
\Big(\m\frac{\p}{\p\m}+\g_{E}\,(t\frac{\p}{\p t}+1)+2\g_{\phi}\Big)\overline{<\phi(x,t)\phi^{\dagger}(0,0)>_{\text{conn.}}}=0\,.
\ee
We can find the most general solution to the above equation. It is determined up to an unknown function and is given by
\be
\overline{<\phi(x,t)\phi^{\dagger}(0,0)>_{\text{conn.}}}=\frac{\m^{-(2\g_\phi+\g_E)}}{|\vec x|^{2\Delta+2\g_\phi+\g_E}}f\Big(\frac{t}{\m^{\g_E}|\vec x|^{2+\g_E}},g_{*},\lambda_{*}\Big)\,,
\ee
where $\Delta$ is the classical scaling dimension.
We see from the above that the two point function is invariant under the scale transformation $\phi\rightarrow \alpha^{-\Delta*}\phi$, $t\rightarrow \alpha^{2+\g_{E}}t$ and $|\vec x|\rightarrow \alpha |\vec x|$ with $\Delta_{*}=\Delta+\g_{\phi}+\frac{\g_{E}}{2}$.
\subsection{Deformation by fermion bilinears}
We can repeat the above analysis for the deformations due to the fermion bilinears, i.e. $\Psi^{\dagger}\Psi$ and $\Psi^{\dagger}\sigma_{3}\Psi$. In order to treat these deformations uniformly, we consider the bilinear $\Psi^{\dagger}\Lambda\Psi$, where $\Lambda$ is either $\mathbb{I}_{2\times2}$ or $\sigma_{3}$. In this case, the replicated action is
\bea
S^{(f)}_{\text{repl.}}=\sum_{A=1}^{\underline {n}}S_{0,A}+i\frac{v}{2}\int d^{d}x\,dt\,dt'\,\sum_{A,B=1}^{\underline {n}}\Big(\Psi_{A}^{\dagger}(x,t)\Lambda\Psi_{A}(x,t)\Big)\Big(\Psi_{B}^{\dagger}(x,t')\Lambda\Psi_{B}(x,t')\Big)\,,\nn\\
\eea
where $S_{0,A}$ is given in~\eqref{Replic.PureAction}. With the above action, we can now perform the one-loop computation. As we see shortly, the calculations are very similar to the case of scalar deformation, 
however, the renormalization flow is very different. 

\begin{figure}[htpb]
\begin{center}
\vspace{-9 cm}
\includegraphics[width=6in]{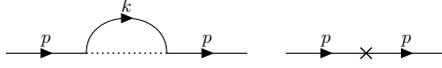}
\vspace{-8 cm}
\caption{Fermion two point function in the presence of the disorder vertex.}\label{fermion2ptDisorder}
\end{center}
\end{figure}

First, we compute the two-point function at one loop. The scalar two-point function receives contribution only due to the fermions in the loop as in the section~\ref{NRCFT}, and gives the wave function renormalization, $Z_{\phi}$~\eqref{DeltaPhiscalar}.
In contrast to the previous analysis, the fermionic two-point function is now non-trivial. The two-point function of the fermions $<T\Psi_{A}(t,x)\Psi^{\dagger}_{B}(t',x')>$ receives a contribution to due to the disorder vertex, as shown in figure~\ref{fermion2ptDisorder}, and gives renormalization to the frequency part of the kinetic term. 
The contributions are
\bea
&&-\delta_{A,B}\lambda\int\frac{dE}{2\pi}\int\prod_{i=1,2}\frac{d^{d}p_{i}}{(2\pi)^{d}}e^{-iE_{1}(t_{1}-t_{2})-i\vec p_{1}\cdot(\vec x_{1}-\vec x_{2})}\Big(G(E_{1},\vec p_{1})G(E_{1},\vec p_{2})G(E_{1},\vec p_{1})\Big)_{\alpha\beta}\nn\\
&&+i\delta_{A,B}\delta_{\alpha}\int\frac{dE}{2\pi}\int\frac{d^{d}p_{1}}{(2\pi)^{d}}e^{-iE_{1}(t_{1}-t_{2})-i\vec p_{1}\cdot(\vec x_{1}-\vec x_{2})}E_{1}\Big(G(E_{1},\vec p_{1})G(E_{1},\vec p_{1})\Big)_{\alpha\beta}\,.
\eea
Cancelling the divergence, we obtain
\be
\delta_{\alpha}=-\frac{\lambda}{2\pi^{2}\e}\,.
\ee

Next, we compute the four-point function. The four-point function of the scalar receives contribution due to the fermions in the loop and, as we have observed previously, it is finite. 
The fermion four-point function, $<T\Psi^{\dagger}_{A_{1}}(t_{1},x_{x})\Psi^{\dagger}_{A_{2}}(t_{2},x_{2})\Psi_{A_{3}}(t_{3},x_{3})\Psi_{A_{4}}(t_{4},x_{4})>$, is divergent and the divergence is due to the disorder vertex. The cancellation of the divergence requires
\be
\delta_{\lambda}=-\frac{2\lambda}{\pi^{2}\e}\,.
\ee 

Let us compare the above results with the scalar case~\eqref{DeltaAlphaScalar} and~\eqref{deltaLambScalar}. 
We note the difference in factor 4. 
The difference can be traced to the square of the particle number. In our convention, the fermion and scalar carry the particle number 1 and 2, respectively.
Furthermore, we also note that there is no wave function renormalization for the fermionic field. It contrasts the scalar case, and, as we will see, it is responsible for a very different renormalization group flow.

Now, with the above results in hand, we can compute the $\beta$-function. These are given by
\bea
&&\beta_{g}=-\frac{\epsilon g}{2}+\frac{g^{3}}{16\pi^{2}}\,,\nn\\
&&\beta_{\lambda}=-\epsilon\lambda-\frac{2\lambda^{2}}{\pi^{2}}\,.
\eea
The above $\beta$-function equations are very different compare to the scalar deformation~\eqref{BetaFnScalar}. One of the major reasons for the difference is the absence of the renormalization of the fermionic field. The physical fixed points are obtained by solving $\beta_{g}=0=\beta_{\lambda}$ and are as follows:
\bea
(\lambda_{*}, g_{*})=(0,0),(0,2\pi\sqrt{2\epsilon{}})\,.
\eea
Thus, in this case, we do not find a non-trivial disorder fixed point.
\section{Disorder system at finite temperature}\label{Finite Temp. phases}
This section aims to study the thermodynamics and phases of the disorder unitary fermion system at finite temperatures. To understand the phase diagram, we need to find the free energy of the system as a function of temperature and chemical potential. The thermodynamics of the unitary fermion system with non-zero chemical potential has been studied in~\cite{PhysRevA.75.063618}. In the paper, the author investigated the phases of the unitary system in the imaginary time formalism in $\e$-expansion when the chemical potential is of the order of $\e$. 
In the present section, we follow their analysis, but in the presence of disorder. 

The basic idea would be to compute the quenched average free energy at a finite temperature with chemical potentials. It is given as
\be
\bar F(T,\m_{i})=-\lim_{\underline {n}\rightarrow 0}\frac{\p}{\p \underline {n}}\int \prod_{A=1}^{\underline {n}}[d\eta_{A}^{\dagger}][d\eta_{A}]\,e^{-S_{\text{repl.}}}=\lim_{\underline {n}\rightarrow 0}\Big(e^{- \Gamma_{\text{repl.}}}\frac{\p \Gamma_{\text{repl.}}}{\p \underline {n}}\Big)\,,
\ee
where $T$ and $\m_{i}$ are temperature and chemical potentials, respectively.
In the above we have defined the replicated effective potential as
\be
e^{- \Gamma_{\text{repl.}}}=\int \prod_{A=1}^{\underline {n}}[d\eta_{A}^{\dagger}][d\eta_{A}]\,e^{-S_{\text{repl.}}}\,.
\ee
As we will see below $\Gamma_{\text{repl.}}$ will depend on the number of replicas and vanishes in the limit $\underline {n}\rightarrow 0$. Our goal would be to study the behaviour of the quenched average free energy as a function of the temperature and chemical potential.

Our analysis starts with the Euclidean action which is
\bea
S&=&\int_{0}^{\beta} d\tau\int d^{d}x\Big[\Psi^{\dagger}(\p_{\tau}-\sigma_{3}\frac{\nabla^{2}}{2})\Psi\Big]+\int_{0}^{\beta} d\tau\int d^{d}x\Big[\phi^{\dagger}(\p_{\tau}-\frac{\nabla^{2}}{4})\phi\Big]\nn\\
&&-\int_{0}^{\beta} d\tau\int d^{d}x\Big[\tilde\m\Psi^{\dagger}\sigma_{3}\Psi-\delta\,\phi^{\dagger}\phi+g_{0}\Psi^{\dagger}\sigma_{+}\Psi \phi+g_{0}\Psi^{\dagger}\sigma_{-}\Psi\phi^{\dagger}\Big]-\int_{0}^{\beta} d\tau\int d^{d}x\,h(x)\phi^{\dagger}\phi\,.\nn\\
\eea
Here we consider only the deformation by the disorder operator $\phi^{\dagger}\phi$.
After integrating over the disorder field, the replicated action is
\be\label{replicatedAction}
S_{\text{repl.}}=\sum_{A=1}^{\underline {n}}S_{0,A}-\frac{v}{2}\int_{0}^{\beta} d\tau\int_{0}^{\beta}d\tau'\,\int d^{d}x\,\sum_{A,B=1}^{\underline {n}}\phi_{A}^{\dagger}(x,\tau)\phi_{A}(x,\tau)\phi_{B}^{\dagger}(x,\tau')\phi_{B}(x,\tau')\,,
\ee
where 
\bea
S_{0,A}&=&\int_{0}^{\beta} d\tau\int d^{d}x\Big[\Psi_{A}^{\dagger}(\p_{\tau}-\sigma_{3}\frac{\nabla^{2}}{2})\Psi_{A}\Big]+\int_{0}^{\beta} d\tau\int d^{d}x\Big[\phi_{A}^{\dagger}(\p_{\tau}-\frac{\nabla^{2}}{4})\phi_{A}\Big]\nn\\
&&-\int d\tau\,d^{d}x\Big[\tilde\m\Psi_{A}^{\dagger}\sigma_{3}\Psi_{A}
-\delta\,\phi_{A}^{\dagger}\phi_{A}+g_{0}\Psi_{A}^{\dagger}\sigma_{+}\Psi_{A} \phi_{A}+g_{0}\Psi_{A}^{\dagger}\sigma_{-}\Psi_{A}\phi_{A}^{\dagger}\Big]\,.\nn\\
\eea
Next, we use the above action to compute the free energy.
\subsection{Effective action for the zero frequency mode}
Before proceeding to compute the free energy, let us comment on the (replicated) action for the zero frequency mode.
To obtain this, we first write down the action in various Matsubara modes and then integrate out all the non-zero frequency modes. The result would be the effective action for the zero-frequency mode. 
The expansion of the fields in terms of Matsubara modes is 
\be
\phi_{A}(x,\tau)=T\sum_{n}e^{-i\omega_{n}\tau}\phi_{A,n}(x),\quad \Psi_{A}(x,\tau)=T\sum_{n}e^{-i\omega^{f}_{n}\tau}\Psi_{A,n}(x)\,,
\ee
where $\omega_{n}=\frac{2\pi n}{\beta}$ and $\omega^{f}_{n}=\frac{2\pi (n+\frac{1}{2})}{\beta}$.
Substituting the above expansion in the action~\eqref{replicatedAction}, we obtain
\bea
S_{\text{repl.}}&=&T\sum_{A}\sum_{n}\int d^{d}x\,\Big[\Psi_{A,n}^{\dagger}(-i\omega_{n}^{f}-\sigma_{3}\frac{\nabla^{2}}{2})\Psi_{A,n}+\phi_{A,n}^{\dagger}(-i\omega_{n}-\frac{\nabla^{2}}{4})\phi_{A,n}\Big]\nn\\
&&-T\sum_{A}\sum_{n}\int d^{d}x\Big[\tilde\m\Psi_{A,n}^{\dagger}\sigma_{3}\Psi_{A,n}
-\delta\,\phi_{A,n}^{\dagger}\phi_{A,n}+g_{0}T\sum_{n'}\Big(\Psi_{A,n'}^{\dagger}\sigma_{+}\Psi_{A,n} \phi_{A,n'-n}\nn\\
&&+\Psi_{A,n'}^{\dagger}\sigma_{-}\Psi_{A,n}\phi_{A,n-n'}^{\dagger}\Big)\Big]-\frac{v\, T^{2}}{2}\int d^{d}x\,\sum_{A,B}\sum_{n,n'}\phi_{A,n}^{\dagger}(x)\phi_{A,n}(x)\phi_{B,n'}^{\dagger}(x)\phi_{B,n'}(x)\,.\nn\\
\eea
We separate the zero frequency mode and place all the terms containing non-zero frequency mode in $S_{n\neq0}$.
Thus, we have
\bea
S_{\text{repl.}}&=&T\sum_{A}\int d^{d}x\,\phi_{A,0}^{\dagger}(-\frac{\nabla^{2}}{4}+\delta)\phi_{A,0}-\frac{v\, T^{2}}{2}\int d^{d}x\,\sum_{A,B}\phi_{A,0}^{\dagger}(x)\phi_{A,0}(x)\phi_{B,0}^{\dagger}(x)\phi_{B,0}(x)\nn\\
&&+S_{n\neq 0}\,,
\eea
and
\bea
S_{n\neq0}&=&T\sum_{A}\sum_{n}\int d^{d}x\,\Psi_{A,n}^{\dagger}(-i\omega_{n}^{f}-\sigma_{3}\frac{\nabla^{2}}{2}-\tilde\m\,\sigma_{3})\Psi_{A,n}\nn\\
&&+T\sum_{A}\sum_{n\neq 0}\int d^{d}x\,\phi_{A,n}^{\dagger}(-i\omega_{n}-\frac{\nabla^{2}}{4}+\delta)\phi_{A,n}\nn\\
&&-g_{0}T^{2}\sum_{A}\sum_{n,n'}\int d^{d}x\Big(\Psi_{A,n'}^{\dagger}\sigma_{+}\Psi_{A,n} \phi_{A,n'-n}+\Psi_{A,n'}^{\dagger}\sigma_{-}\Psi_{A,n}\phi_{A,n-n'}^{\dagger}\Big)\nn\\
&&-\frac{v\, T^{2}}{2}\int d^{d}x\,\sum_{A,B}\sum\limits_{\substack{n,n'\\(n,n')\neq(0,0)}}\phi_{A,n}^{\dagger}(x)\phi_{A,n}(x)\phi_{B,n'}^{\dagger}(x)\phi_{B,n'}(x)\,,\nn\\
&=&T\sum_{A}\sum_{n}\int d^{d}x\,\Psi_{A,n}^{\dagger}(-i\omega_{n}^{f}-\sigma_{3}\frac{\nabla^{2}}{2}-\tilde\m\,\sigma_{3})\Psi_{A,n}\nn\\
&&+T\sum_{A}\sum_{n\neq 0}\int d^{d}x\,\phi_{A,n}^{\dagger}(-i\omega_{n}-\frac{\nabla^{2}}{4}+\delta)\phi_{A,n}
+S_{I}\,.
\eea
The effective action for the zero frequency mode is obtained by performing the integration over all the non-zero modes. It can be done in the perturbation expansion: we expand the exponential involving terms proportional to $g_{0}$ and $v$ and contract the non-zero modes using Wick's contraction. This generates temperature-dependent corrections to the coefficients of all the powers of $\sum_{A}\phi_{A,0}^{\dagger}\phi_{A,0}$. However, each contribution is UV divergent, and $T$-independent counter terms can cure the divergences. Thus, the effective action is
\bea
S^{\text{eff.}}_{0;\text{repl.}}&=&T\sum_{A}\int d^{d}x\,(1+\delta_{\phi})\,\phi_{A,0}^{\dagger}\Big(-\frac{\nabla^{2}}{4}\Big)\phi_{A,0}+T(\delta(T)+\delta r)\sum_{A}\int d^{d}x\,\phi_{A,0}^{\dagger}\phi_{A,0}\nn\\
&&-\frac{(v(T)+\delta v) T^{2}}{2}\int d^{d}x\,\sum_{A,B}\phi_{A,0}^{\dagger}(x)\phi_{A,0}(x)\phi_{B,0}^{\dagger}(x)\phi_{B,0}(x)+....\,,
\eea
where $\delta_{\phi},\delta r$ and $\delta v$ are temperature independent divergences. 
Note that the action for the zero mode is that of $U(\underline n)$-scalar field in $d$-dimensions. 

We can calculate the first few corrections to the effective action. These are computed by integrating out the non-zero modes as
\be
\mathcal C\int\prod_{A}d\phi_{A,0}\, d\phi^{\dagger}_{A,0}\, e^{-S_{0;\text{repl.}}}[1-<S_{I}>+\frac{1}{2}<S_{I}^{2}>+...]\,,
\ee
where the notation ``$<...>$'' denotes the expectation value computed in the non-zero modes. 
Now, each of these expectation values can be calculated using Wick's contractions of non-zero modes. For example, the first two terms are~\footnote{We have absorbed the contribution of the bubble diagrams in the definition of $\mathcal C$.}
\bea
<S_{I}>&=&\underline {n}\,v T\int d^{d}x\sum_{A}\phi_{A,0}^{\dagger}(x)\phi_{A,0}(x)\int\frac{d^{d}p}{(2\pi)^{d}}\sum_{n\neq0}\frac{1}{i\omega_{n}-(\frac{\e_{p}}{2}+\delta)},\nn\\
&=&\underline {n}\,v T\int d^{d}x\sum_{A}\phi_{A,0}^{\dagger}(x)\phi_{A,0}(x)\int\frac{d^{d}p}{(2\pi)^{d}}\Big(\frac{1}{\frac{\e_{p}}{2}+\delta}-\frac{\beta}{2}(\coth\frac{\beta(\frac{\e_{p}}{2}+\delta)}{2}-1)\Big)\,,\nn\\
\eea
and
\bea
<S^{2}_{I}>&=&2g_{0}^{2}T\sum_{A}\int \frac{d^{d}p}{(2\pi)^{d}}\phi^{\dagger}_{A,0}(p)\phi_{A,0}(p)f(p)+<S_{I}>^{2}\nn\\
&&+\underline {n}\,v^{2}T\int\prod_{i=1,2,3}\frac{d^{d}k_{i}}{(2\pi)^{d}}\sum_{A,B}\phi_{A,0}^{\dagger}(k_{1}+k_{3}-k_{2})\phi_{A,0}(k_{1})\phi^{\dagger}_{B,0}(k_{2})\phi_{B,0}(k_{3})\tilde f(k_{3}-k_{2})\,,\nn\\
&&+2\underline {n}\,v^{2}T\int\prod_{i=1}^{2}\frac{d^{d}k_{i}}{(2\pi)^{d}}\tilde g(k_{1},k_{2})\sum_{A}\int \frac{d^{d}p}{(2\pi)^{d}}\phi^{\dagger}_{A,0}(p)\phi_{A,0}(p)\nn\\
&&+2\underline {n}^{2}\,v^{2}T\int\prod_{i=1}^{2}\frac{d^{d}k_{i}}{(2\pi)^{d}}g(k_{1},k_{2})\sum_{A}\int \frac{d^{d}p}{(2\pi)^{d}}\phi^{\dagger}_{A,0}(p)\phi_{A,0}(p)\,.\nn\\
\eea
The explicit expression for the functions appearing in the above integrals are 
\bea
&&f(p)=\int\frac{d^{d}k}{(2\pi)^{d}}\frac{1}{\epsilon_{\vec k}+\epsilon_{\vec k+\vec p}-2\tilde\m}\Big[1-\frac{1}{e^{\beta(\epsilon_{\vec k}-\tilde\m)}+1}-\frac{1}{e^{\beta(\epsilon_{\vec k+\vec p}-\tilde\m)}+1}\Big]\,,\nn\\
&&\tilde f(p)=-\int\frac{d^{d}k}{(2\pi)^{d}}\frac{2}{|\epsilon_{\vec k+\vec p}-\epsilon_{\vec k}|}\Big[\frac{1}{e^{-\beta(\tilde C-E)}-1}-\frac{1}{e^{-\beta(\tilde C+E)}-1}\Big]-\int\frac{d^{d}k}{(2\pi)^{d}}\frac{4T}{(\epsilon_{\vec k}+2\delta)(\epsilon_{\vec k+\vec p}+2\delta)}\,,\nn\\
&&\tilde C=-\frac{1}{4}(\epsilon_{\vec p+\vec k}+\epsilon_{\vec k}+2\delta),\quad E=\frac{1}{4}|\epsilon_{\vec p+\vec k}-\epsilon_{\vec k}|\,,\nn\\
&&\tilde g(k_{1},k_{2})=\frac{8}{(\e_{k_{1}}+2\delta)^{2}(\e_{k_{2}}+2\delta)}-\frac{\beta^{2}}{2(\e_{k_{2}}-\e_{k_{1}})}\frac{1}{\sinh^{2}\frac{\beta}{2}(\frac{\e_{k_{1}}}{2}+\delta)}\nn\\
&&\qquad\qquad-\frac{4\beta}{(\e_{k_{2}}-\e_{k_{1}})^{2}}\Big(\frac{1}{e^{\beta(\frac{\e_{k_{3}}}{2}+\delta)}-1}-\frac{1}{e^{\beta(\frac{\e_{k_{1}}}{2}+\delta)}-1}\Big)\,,\nn\\
&&g(k_{1},k_{2})=-\Big(\frac{1}{(\frac{\e_{k_{1}}}{2}+\delta)^{2}}-\frac{\beta^{2}}{4\sinh^{2}\frac{\beta}{2}(\frac{\e_{k_{1}}}{2}+\delta)}\Big)\Big(\frac{1}{\frac{\e_{k_{2}}}{2}+\delta}-\frac{\beta}{2}(\coth\frac{\beta(\frac{\e_{k_{2}}}{2}+\delta)}{2}-1)\Big)\,.\nn\\
\eea
These are obtained using Wick's contraction:
\bea
&&\wick{\c1\phi_{A}(i\omega_{n_{1}},x_{1})\c1\phi_{B}^{\dagger}(i\omega_{n_{2}},x_{2})}=\beta\delta_{A,B}\delta_{n_{1},n_{2}}\int\frac{d^{d}k}{(2\pi)^{d}}\frac{-e^{-i\vec k\cdot(\vec x_{1}-\vec x_{2})}}{i\omega_{n_{1}}-(\frac{\e_{k}}{2}+\delta)}\,,\nn\\
&&\wick{\c1\Psi_{A}(i\omega_{n_{1}},x_{1})\c1\Psi_{B}^{\dagger}(i\omega_{n_{2}},x_{2})}=\beta\delta_{A,B}\delta_{n_{1},n_{2}}\int\frac{d^{d}k}{(2\pi)^{d}}\frac{-e^{-i\vec k\cdot(\vec x_{1}-\vec x_{2})}}{(i\omega^{f}_{n_{1}})^{2}-(\e_{k}-\tilde\m)^{2}}\nn\\&&
\qquad\qquad\qquad\qquad\qquad\qquad\qquad\qquad\times\begin{pmatrix}i\omega^{f}_{n_{1}}+(\e_{k}-\tilde\m)&0\\0&i\omega^{f}_{n_{1}}-(\e_{k}-\tilde\m)\end{pmatrix}\,.\nn\\
\eea
One finds the terms $\delta_{\phi},\delta v$ and $\delta r$ by looking at the divergences in the above expressions. At this order (upto one loop) in the computation, these are
\bea
\delta_{\phi}=\frac{g_{0}^{2}}{8\pi^{2}\e},\quad \delta r=-\frac{g_{0}^{2}\tilde\m}{4\pi^{2}\e}-{\underline n}\frac{2v\delta}{\pi^{2}\e},\quad\delta v=-{\underline n}\frac{2v^{2}}{\pi^{2}\e}\,.
\eea
The above divergences can be removed by introducing counter terms. However, these divergences, in particular $\delta v$, reveal something interesting. It shows that the running of the disorder coupling constant, as observed in section~\ref{DisorderOperatorPhi}, is primarily due to the Wilsonian RG flow in the zero mode sector. The contribution of the non-zero modes to the $\beta$-function is proportional to ${\underline n}$ and disappears from the generalized Callan-Symanzik equation.
\subsection{Effective potential for constant mode near $4$-spatial dimensions}
In the present section, we compute the one-loop effective potential for the scalar in $(4-\e)$-dimensions. For the computation, we assume that the coupling constant $g$ and $\lambda$ do not scale with $\e$. In the next section, we will relax this assumption.
We expand the action about the replica symmetric vacuum.
\be
\phi_{A}=\phi_{0}+\hat\phi_{A},\quad \phi^{\dagger}_{A}=\phi_{0}^{\dagger}+\hat\phi^{\dagger}_{A}\,.
\ee
We are interested in the effective potential for the field $\phi_{0}$ and $\phi_{0}^{\dagger}$, and, therefore, we assume that these are independent of space and time coordinates. Note that the assumption is consistent with the fact that the replicated theory respects the translational invariance of space and time.

We first integrate out the fermions. The relevant terms in the action are
\be
\int_{0}^{\beta} d\tau\int d^{d}x\Big[\Psi_{A}^{\dagger}(\p_{\tau}-\sigma_{3}\frac{\nabla^{2}}{2})\Psi_{A}\Big]
-\int d\tau\,d^{d}x\Big[\tilde\m\Psi_{A}^{\dagger}\sigma_{3}\Psi_{A}
+g\Psi_{A}^{\dagger}\sigma_{+}\Psi_{A} \phi_{0}+g\Psi_{A}^{\dagger}\sigma_{-}\Psi_{A}\phi_{0}^{\dagger}\Big]\,.
\ee
The one loop contribution of fermions to the 1PI (replicated) effective potential is ($\Gamma_{\text{repl.}}=\beta\text{Vol.}V_{\text{repl.}}$)
\be
V^{f}_{\text{repl.}}=-\underline {n} \,T\sum_{\omega^{f}_{m}}\int_{\vec p} \ln \Big[(\omega^{f}_{m})^{2}+E_{p}^{2}\Big]=-\underline {n}\int_{\vec p}\Big[E_{p}+2T\ln(1+e^{-\frac{E_{p}}{T}})\Big]\,,
\ee
where $E_{p}^{2}=(\epsilon_{\vec p}-\tilde\m)^{2}+g_{0}^{2}\phi_{0}\phi_{0}^{\dagger},\,\,\omega^{f}_{m}=2\pi(m+\frac{1}{2})T$ and $\text{Vol.}$ is the spatial volume. Also, in the above we have used the notation $\int_{\vec p}$ to indicate the integration over spatial momentum, i.e. $\int_{\vec p}=\int \frac{d^{d}p}{(2\pi)^{d}}$.

Next, we look at the contribution from the scalar fluctuations. The quadratic part of the action which is relevant for our computation is (we have ignored the term linear in $\hat \phi_{A}$, since it does not contribute to 1PI effective action)
\bea
&&\int d\tau\,d^{d}x\sum_{A}\Big[\hat\phi_{A}^{\dagger}(\p_{\tau}-\frac{\nabla^{2}}{4})\hat\phi_{A}\Big]+\int d\tau\,d^{d}x\sum_{A}\Big[\delta (\phi_{0}^{\dagger}\phi_{0}+\hat\phi_{A}^{\dagger}\hat\phi_{A})\Big]\nn\\
&&-\frac{\lambda}{2}\int d^{d}x\,d\tau\,d\tau'\,\sum_{A,B=1}^{\underline n}\Big[(\phi_{0}^{\dagger}\phi_{0})^{2}+2\phi_{0}^{\dagger}\phi_{0}\Big(\hat\phi_{A}^{\dagger}(x,\tau)\hat\phi_{A}(x,\tau)+\hat\phi_{A}(x,\tau)\hat\phi_{B}^{\dagger}(x,\tau')\Big)\nn\\
&&\qquad\qquad \qquad\qquad+(\phi_{0}^{\dagger})^{2}\hat\phi_{A}(x,\tau)\hat\phi_{B}(x,\tau')+(\phi_{0})^{2}\hat\phi_{A}^{\dagger}(x,\tau)\hat\phi^{\dagger}_{B}(x,\tau')\Big]\,.\nn\\
\eea
Thus, the contribution of the scalar to the 1PI effective potential at one loop is
\bea
V^{s}_{\text{repl.}}&=&\underline {n}\,\delta\,\phi_{0}^{\dagger}\phi_{0}-\frac{\lambda}{2T}\underline {n}^{2}(\phi_{0}^{\dagger}\phi_{0})^{2}+T\sum_{\omega_{m}}\int_{\vec p}\Big[\frac{\underline {n}}{2}\ln\Big(\omega_{m}^{2}+\Big(\frac{\e_{\vec p}}{2}+\delta-\frac{\underline {n}\,\lambda}{T}\phi_{0}^{\dagger}\phi_{0}\Big)^{2}\Big)\Big]\nn\\
&&+\frac{T}{2}\int_{\vec p}\ln\frac{\frac{\e_{\vec p}}{2}+\delta-\frac{3\underline {n}\,\lambda}{T}\phi_{0}^{\dagger}\phi_{0}}{\frac{\e_{\vec p}}{2}+\delta-\frac{\underline {n}\,\lambda}{T}\phi_{0}^{\dagger}\phi_{0}}\,.
\eea
Note that the first two terms are classical contributions. Summing over the frequencies, we obtain
\bea
V^{s}_{\text{repl.}}&=&\underline {n}\,\delta\,\phi_{0}^{\dagger}\phi_{0}-\frac{\lambda}{2T}\underline {n}^{2}(\phi_{0}^{\dagger}\phi_{0})+\underline {n}\int_{\vec p} \Big[\frac{1}{2}(\frac{\e_{\vec p}}{2}+\delta-\frac{\underline {n}\,\lambda}{T}\phi_{0}^{\dagger}\phi_{0}\Big)+T\ln\Big(1-e^{-\frac{\frac{\e_{\vec p}}{2}+\delta-\frac{\underline {n}\,\lambda}{T}\phi_{0}^{\dagger}\phi_{0}}{T}}\Big)\Big]\nn\\
&&+\frac{T}{2}\int_{\vec p}\ln\frac{\frac{\e_{\vec p}}{2}+\delta-\frac{3\underline {n}\,\lambda}{T}\phi_{0}^{\dagger}\phi_{0}}{\frac{\e_{\vec p}}{2}+\delta-\frac{\underline {n}\,\lambda}{T}\phi_{0}^{\dagger}\phi_{0}}\,.
\eea
Finally, the quenched average effective potential at one loop ($\bar V=\bar F$) is
\bea\label{EffectivePot.}
\bar V(T,\m_{i},\phi_{0})&=&\delta\,\phi_{0}^{\dagger}\phi_{0}-\int_{\vec p}\Big[E_{p}+2T\ln(1+e^{-\frac{E_{p}}{T}})\Big]+\int_{\vec p} \Big[\frac{1}{2}(\frac{\e_{\vec p}}{2}+\delta\Big)+T\ln\Big(1-e^{-\frac{\frac{\e_{\vec p}}{2}+\delta}{T}}\Big)\Big]\nn\\
&&-\lambda\phi_{0}^{\dagger}\phi_{0}\int_{\vec p}\frac{1}{\frac{\e_{\vec p}}{2}+\delta}\,.
\eea
Note that the third term is a constant, i.e. independent of $\phi_{0}^{\dagger}\phi_{0}$.

\subsubsection{Renormalization of the effective potential}
The effective potential~\eqref{EffectivePot.} suffers from UV divergence in $d=4$-spatial dimension; however, the divergence comes from the $T=0$ part of the integration. We can remove these divergences by adding counter terms $\delta r$ and $\Lambda$. With these, the effective potential is
\bea
\bar V(T,\m_{i},\phi_{0})&=&(\delta+\delta r)\,\phi_{0}^{\dagger}\phi_{0}-\int_{\vec p}\Big[E_{p}+2T\ln(1+e^{-\frac{E_{p}}{T}})\Big]+\int_{\vec p} \Big[\frac{1}{2}(\frac{\e_{\vec p}}{2}+\delta\Big)+T\ln\Big(1-e^{-\frac{\frac{\e_{\vec p}}{2}+\delta}{T}}\Big)\Big]\nn\\
&&-\lambda\phi_{0}^{\dagger}\phi_{0}\int_{\vec p}\frac{1}{\frac{\e_{\vec p}}{2}+\delta}+\Lambda\,.
\eea
We choose $\delta_{r}$ such that at $T=0$, there are no divergence in the $\frac{\p^{2}\bar V}{\p\phi\p\phi^{\dagger}}\Big|_{\phi=0}$. Similarly, we determine $\Lambda$ by requiring that $\bar V(T,\m_{i},\phi_{0}=0)=0$. 
The first condition implies that
\be
\delta r-\frac{g^{2}\tilde\m}{4\pi^{2}\e}+\frac{2\delta\,\lambda}{\pi^{2}\e}=0,\quad\Rightarrow\quad \delta r=\frac{g^{2}\tilde\m}{4\pi^{2}\e}-\frac{2\delta\lambda}{\pi^{2}\e}\,.
\ee
The second renormalization condition implies that
\be
\Lambda=\int_{\vec p}|(\e_{p}-\tilde\m)|-\int_{\vec p} \frac{1}{2}\Big(\frac{\e_{\vec p}}{2}+\delta\Big)
\ee
With these renormalizations, the effective potential is finite and is given by
\bea
\bar V(T,\m_{i},\phi_{0})&=&(\delta+\delta r)\,\phi_{0}^{\dagger}\phi_{0}-\int_{\vec p}\Big[E_{p}-|(\e_{p}-\tilde\m)|+2T\ln(1+e^{-\frac{E_{p}}{T}})\Big]+T\int_{\vec p} \ln\Big(1-e^{-\frac{\frac{\e_{\vec p}}{2}-\delta}{T}}\Big)\nn\\
&&-\lambda\phi_{0}^{\dagger}\phi_{0}\int_{\vec p}\frac{1}{\frac{\e_{\vec p}}{2}+\delta}\,.
\eea
One can analyse the above potential for various values of $T$. For a fixed value of $T$, the potential is a function of the chemical potentials and $\phi_{0}$. 
\subsubsection{Disorder system at $T=0$}
We look at the effective potential for $T=0$. In this case, the effective potential is
\be
\bar V(0,\m_{i},\phi_{0})=(\delta+\delta r)\,\phi_{0}^{\dagger}\phi_{0}-\int_{\vec p}\Big[\sqrt{(\epsilon_{\vec p}-\tilde\m)^{2}+g^{2}\phi_{0}\phi_{0}^{\dagger}}-|(\e_{p}-\tilde\m)|\Big]-\lambda\phi_{0}^{\dagger}\phi_{0}\int_{\vec p}\frac{1}{\frac{\e_{\vec p}}{2}+\delta}\,.
\ee
In the above, we have ignored the $\phi_{0}$-independent terms.
Next, we want to see if there is a non-trivial minimum of the effective potential as a function of $\phi_{0}$.  For this, we need to calculate the integral 
\bea
f(|\phi_{0}|)&=&\int\frac{d^{d}p}{(2\pi)^{d}}\frac{g^{2}|\phi_{0}|}{\sqrt{(\epsilon_{\vec p}-\tilde\m)^{2}+g^{2}|\phi_{0}|^{2}}}\nn\\
&=&\frac{g^{2}|\phi_{0}|\tilde\m^{\frac{d}{2}-1}S_{d-1}}{(2\pi)^{d}}\int_{0}^{\infty}q^{d-1}dq\frac{1}{\sqrt{(\frac{q^{2}}{2}-1)^{2}+\tilde a^{2}}}\,.
\eea
Here $S_{d-1}=\frac{2\pi^{\frac{d}{2}}}{\Gamma(\frac{d}{2})}$ and $\tilde a=\frac{g|\phi_{0}|}{\tilde\m}$. Also, we have assumed that $\tilde\m>0$.
The above integration can be performed explicitly~\eqref{EffectivePot}. We find that for $d=(4-\e)$ and to the leading order in $\e$,
\bea
f(|\phi_{0}|)&=&\frac{g^{2}\tilde\m|\phi_{0}|}{2\pi^{2}\e}+\frac{g^{2}\tilde\m|\phi_{0}|}{4\pi^{2}}\Big(\ln\frac{4\pi}{g|\phi_{0}|}-\g\Big)\nn\\
&&-\frac{g^{2}\tilde\m|\phi_{0}|}{4\pi^{2}}\Big(\sqrt{1+\tilde a^{2}}-\frac{1}{2}\ln\frac{\sqrt{1+\tilde a^{2}}+1}{\sqrt{1+\tilde a^{2}}-1}\Big)\,,
\eea
where $\g$ is Euler's constant.
Let us evaluate the above expression for $\tilde\m<<g|\phi_{0}|$ and $\tilde\m>>g|\phi_{0}|$. In the first case, we have
\be
f(|\phi_{0}|)=\frac{g^{2}|\phi_{0}|}{4\pi^{2}}\Big[-(\g\tilde\m+g|\phi_{0}|+\tilde\m\ln\frac{g|\phi_{0}|}{4\pi})+\frac{2\tilde\m}{\epsilon}\Big]\sim\frac{g^{2}|\phi_{0}|}{4\pi^{2}}\Big[-g|\phi_{0}|+\frac{2\tilde\m}{\epsilon}\Big]\,.
\ee
Thus the first derivative of the potential is
\be
\bar V(0,\m_{i},\phi_{0})'\sim 2(\delta+\delta r)\,|\phi_{0}|-\frac{g^{2}|\phi_{0}|}{4\pi^{2}}\Big[-g|\phi_{0}|+\frac{2\tilde\m}{\epsilon}\Big]+\lambda|\phi_{0}|\Big(\frac{4\delta}{\pi^{2}\e}-\frac{2\delta(\g-1+\ln\frac{\delta}{4\pi})}{\pi^{2}}\Big)\,.
\ee
The counter terms $\delta r$ cancels the $\e^{-1}$-divergences, and thus, we get
\be
\bar V(0,\m_{i},\phi_{0})'\sim 2\delta\,|\phi_{0}|+\frac{g^{3}|\phi_{0}|^{2}}{4\pi^{2}}-2\lambda|\phi_{0}|\frac{\delta(\g-1+\ln\frac{\delta}{4\pi})}{\pi^{2}}=2\delta'\,|\phi_{0}|+\frac{g^{3}|\phi_{0}|^{2}}{4\pi^{2}}\,,
\ee
where $\delta'=\delta(1-\frac{\lambda}{\pi^{2}}(\g-1+\ln\frac{\delta}{4\pi}))$.
We see from the above that there is a non-trivial minima of the potential for
\be\label{minima1}
|\phi_{0}|\sim-\frac{8\pi^{2}\delta'}{g^{3}}\,,
\ee
provided $\lambda(\g-1+\ln\frac{\delta}{4\pi})>\pi^{2}$.

Next, we look at the potential in the region $\tilde\m>>g|\phi_{0}|$. In this case, the function $f(\phi_{0})$ to the leading order in $\frac{g|\phi_{0}|}{\tilde\m}$ is 
\be
f(\phi_{0})=\frac{g^{2}|\phi_{0}|\tilde\m}{2\pi^{2}\e}-\frac{g^{2}|\phi_{0}|\tilde\m}{4\pi^{2}}\Big(1+\g-\ln\frac{8\pi\tilde\m}{g^{2}|\phi_{0}|^{2}}\Big)\,.
\ee
and the derivative of the potential is
\bea
\bar V(0,\m_{i},|\phi_{0}|)'=2\delta'|\phi_{0}|+g^{2}|\phi_{0}|\frac{\tilde\m}{4\pi^{2}}(1+\g-\ln\frac{8\pi\tilde\m}{g^{2}|\phi_{0}|^{2}})\,.
\eea
In this case, the solution seems to be at
\be\label{minima2}
\ln\frac{g|\phi_{0}|}{2\sqrt{2\pi\tilde\m}}=-\frac{1}{2}(1+\g+\frac{8\pi^{2}\delta'}{g^{2}\m})\,.
\ee
\subsection{Effective potential in $\epsilon$-expansion}
In this section, we will discuss the effective potential in $(4-\e)$-spatial dimensions when the coupling constants $g\sim\sqrt{\e}$ and $\lambda\sim\e$ and the background field $\phi_{0}\sim\mathcal O(1)$. This is clearly the case near the critical point~\eqref{CriticalPt}. As is shown in~\cite{Nishida:2006br, PhysRevA.75.063618}, to the leading order in $\e$, the contribution from the fermionic path integration is~\footnote{It is important to emphasize that integrating over fermions generates corrections to the various coefficients of $(\hat\phi^\dagger\hat\phi)^n$ in the scalar action. However, as we have seen in the previous sections, except for $n=1$, the rest of the corrections will be higher powers in $\e$. For example, the correction to the term $(\hat\phi^\dagger\hat\phi)^2$ due to the fermionic loop is proportional to $g^4\sim\e^2$. For $n=1$, we have $\mathcal O(1)$ corrections to the kinetic and chemical potential terms of the scalar. The chemical potential, $\delta$, will be corrected as $\delta-\frac{g^{2}\tilde\m}{4\pi^{2}\e}$. We will rescale the scalar field for all the analyses below to have the standard kinetic term. It results in the rescaling of the coupling constant. However, not to clutter the notation, we will keep using the same notation for the chemical potential and coupling constant. }
\be
V^{f}_{\text{repl.}}=-\underline {n} \,T\sum_{\omega^{f}_{m}}\int_{\vec p} \ln \Big[(\omega^{f}_{m})^{2}+E_{p}^{2}\Big]=-\underline {n}\int_{\vec p}\Big[E_{p}+2T\ln(1+e^{-\frac{E_{p}}{T}})\Big]\,.
\ee
Next, we look at the scalar integration and compute the contribution to the effective action to the leading order in $\e$. 
Our analysis starts with the Euclidean action 
\bea
S_{\text{repl.}}&=&\sum_{A}\int_{0}^{\beta} d\tau\int d^{d}x\Big[\phi_{A}^{\dagger}(\p_{\tau}-\frac{\nabla^{2}}{4}+\delta)\phi_{A}\Big]\nn\\
&&-\frac{\lambda}{2}\int_{0}^{\beta} d\tau\int_{0}^{\beta}d\tau'\,\int d^{d}x\,\sum_{A,B=1}^{n}\phi_{A}^{\dagger}(x,\tau)\phi_{A}(x,\tau)\phi_{B}^{\dagger}(x,\tau')\phi_{B}(x,\tau')\,,
\eea
and substituting $\phi_{A}=\phi_{0}+\hat\phi_{A}$, we obtain
\bea
S_{\text{repl.}}&=&\sum_{A}\int_{0}^{\beta} d\tau\int d^{d}x\Big[\hat\phi_{A}^{\dagger}(\p_{\tau}-\frac{\nabla^{2}}{4}+\delta)\hat\phi_{A}\Big]+\underline {n}\beta V_{d}\delta\, \phi_{0}\phi_{0}^{\dagger}\nn\\
&&-\frac{\lambda}{2}\int_{0}^{\beta} d\tau\int_{0}^{\beta}d\tau'\,\int d^{d}x\,\sum_{A,B=1}^{N}\Big((\phi^{\dagger}_{0}\phi_{0})^{2}+(\phi^{\dagger}_{0})^{2}\hat\phi_{A}(x,\tau)\hat\phi_{B}(x,\tau')+2\phi_{0}\phi_{0}^{\dagger}\hat\phi_{A}(x,\tau)\hat\phi^{\dagger}_{A}(x,\tau)\nn\\
&&+2\phi_{0}\phi_{0}^{\dagger}\hat\phi_{A}(x,\tau)\hat\phi^{\dagger}_{B}(x,\tau')+2\phi^{\dagger}_{0} \hat\phi_{A}^{\dagger}(x,\tau)\hat\phi_{A}(x,\tau)\hat\phi_{B}(x,\tau')+\phi_{0}^{2}\phi_{A}^{\dagger}(x,\tau)\hat\phi_{B}^{\dagger}(x,\tau')\nn\\
&&+2\phi_{0}\hat\phi_{A}(x,\tau)\hat\phi^{\dagger}_{A}(x,\tau)\hat\phi^{\dagger}_{B}(x,\tau')+\hat\phi_{A}^{\dagger}(x,\tau)\hat\phi_{A}(x,\tau)\hat\phi_{B}^{\dagger}(x,\tau')\hat\phi_{B}(x,\tau')\Big)\,.
\eea
In the above we have ignored terms which are linear in $\hat\phi$ as these would not give 1PI diagrams.

In order to organize the computation, we label the vertices as follows:
\bea
&&(1)=\frac{\lambda}{2}(\phi^{\dagger}_{0})^{2}\hat\phi_{A}(x,\tau)\hat\phi_{B}(x,\tau'),\quad (2)=\lambda {\underline n}\,\phi_{0}\phi_{0}^{\dagger}\hat\phi_{A}(x,\tau)\hat\phi^{\dagger}_{A}(x,\tau),\nn\\
&& (3)=\lambda\phi_{0}\phi_{0}^{\dagger}\hat\phi_{A}(x,\tau)\hat\phi^{\dagger}_{B}(x,\tau')\,,\quad(4)=\lambda\phi^{\dagger}_{0} \hat\phi_{A}^{\dagger}(x,\tau)\hat\phi_{A}(x,\tau)\hat\phi_{B}(x,\tau'),\nn\\
&&(5)=\frac{\lambda}{2} \phi_{0}^{2}\phi_{A}^{\dagger}(x,\tau)\hat\phi_{B}^{\dagger}(x,\tau')\,,\quad(6)=\lambda\phi_{0}\hat\phi_{A}(x,\tau)\hat\phi^{\dagger}_{A}(x,\tau)\hat\phi^{\dagger}_{B}(x,\tau'),\nn\\
&&(7)=\frac{\lambda}{2}\hat\phi_{A}^{\dagger}(x,\tau)\hat\phi_{A}(x,\tau)\hat\phi_{B}^{\dagger}(x,\tau')\hat\phi_{B}(x,\tau')\,.
\eea

At the first order in the perturbative expansion, we only have the contribution from the vertex (3) and is
\be\label{firstorder}
\frac{\lambda}{2}\int_{0}^{\beta} d\tau\int_{0}^{\beta} d\tau'\int d^{d}x\sum_{A,B}2\phi_{0}\phi^{\dagger}_{0}<\hat\phi_{A}(x,\tau)\hat\phi_{B}^{\dagger}(x,\tau')>=\lambda\, {\underline n}\,\phi_{0}\phi_{0}^{\dagger}\beta V_{d}\int\frac{d^{d}k}{(2\pi)^{d}}\frac{1}{\frac{\e_{k}}{2}+\delta}\,.
\ee
In the above, we have used the propagator
\be
<\phi(x,\tau)\phi^{\dagger}(x',\tau')>=T\sum_{n}\int\frac{d^{d}k}{(2\pi)^{d}}e^{-i\omega_{n}(\tau-\tau')}e^{-ik(x-x')}G(i\omega_{n},k)\,,
\ee
with
\be
G(i\omega_{n},k)=\frac{-1}{i\omega_{n}-(\frac{\e_{k}}{2}+\delta)}\,.
\ee
The integral in~\eqref{firstorder} is divergent as $\frac{1}{\e}$ and, therefore, the total contribution is $\mathcal O(1)$ in $\e$-expansion (since $\lambda\sim\e$)\,.
Next, we look at the second-order contribution. Some of the diagrams are of order ${\underline n}^{2}$, and we will ignore these diagrams since these will not contribute to the quenched average effective potential. For example, the diagram involving the vertices $(1)\&(5)$ and $(2)\&(3)$.  We list the diagrams which give the non-trivial contributions: First, consider the diagram involving vertices $(4)\&(6)$. Its contribution is
\bea
\underline {n}\lambda^{2}\phi^{\dagger}_{0}\phi_{0}\beta V_{d}\int \prod_{i=1}^{3}\frac{d^{d}k_{i}}{(2\pi)^{d}}(2\pi)^{d}\delta(k_{1}+k_{2}-k_{3})\prod_{i=1}^{3}\frac{1}{\frac{\e_{k_{i}}}{2}+\delta}\,.
\eea
Now, evaluating the above integral reveals that the integral is divergent and has poles of order $\frac{1}{\e^{2}}$. However, since the integral is multiplying $\lambda^{2}$, the final answer is $\mathcal O(1)$.
Next, consider the diagram involving the vertices $(3)\&(7)$. In this case, the result is
\bea
&&\lambda^{2}\phi_{0}^{\dagger}\phi_{0}\underline {n}\beta\,V_{d}\int\prod_{i=1}^{3}\frac{d^{d}k_{i}}{(2\pi)^{d}}(2\pi)^{d}\delta(k_{1}-k_{2})\prod_{i=1}^{3}\frac{1}{\frac{\e_{k_{i}}}{2}+\delta}\nn\\
&&\quad\quad\qquad\qquad\qquad\qquad=\lambda^{2}\phi_{0}^{\dagger}\phi_{0}\underline {n}\beta\,V_{d}\int\frac{d^{d}k_{1}}{(2\pi)^{d}}\frac{d^{d}k_{3}}{(2\pi)^{d}}\frac{1}{(\frac{\e_{k_{1}}}{2}+\delta)^{2}}\frac{1}{\frac{\e_{k_{3}}}{2}+\delta}\,.
\eea
Note again that the above integral is divergent and the divergence goes like $\frac{1}{\e^{2}}$, but since the integral is multiplying $\lambda^{2}$, therefore, the final result is $\mathcal O(1)$. 

We want to generalize the above observations to higher orders in the loop diagram. It turns out that the only diagrams we need to consider, i.e. which are $\mathcal O(\underline {n})$ and $\mathcal O(1)$ in $\e$, are the ones which have a single vertex $(3)$ but an arbitrary number of vertex $(7)$, and two cubic vertices, $(4)$ and $(6)$, with an arbitrary number of quartic vertex $(7)$. 
This we can see as follows: We scale the scalar field as $\hat\phi\rightarrow \e^{-\frac{1}{4}}\hat\phi$. Using $\lambda\sim\e$, we have the quadratic vertices, cubic vertices and the quartic vertex of the order $\mathcal O(\sqrt{\e}),\mathcal O(\e^{\frac{3}{4}})$ and $\mathcal O(1)$, respectively. Thus, for a given 1PI diagram, which is $\mathcal O(\underline{n})$, with $V_{2}$, $V_{3}$ and $V_{4}$ being the number of quadratic, cubic and quartic vertices, respectively, the power of $\e$ is $\e^{V_{2}+V_{3}+V_{4}}$. Note that we have included the contribution due to the propagator, which is $\e^{-1}$. Next, we need to include contributions from the loop integral. Assuming that a divergent integral in $d=(4-\e)$-spatial dimensions has a simple pole in $\e$, the maximum power of $\e$ associated is $\e^{V_{2}+V_{3}+V_{4}-1}$. Note that this does not give the right power of $\e$ if the total integral has a factorised form. In such cases, if each of these integrals is divergent, the power of $\e$ will be lower than $V_{2}+V_{3}+V_{4}-1$. Now, we note that a 1PI diagram with $L$-loops, where $L=V_{4}+\frac{V_{3}}{2}+1$, can atmost have divergences$\sim\e^{-L}$. Thus, minimal power $\e$ associated with the 1PI diagram is $V_{2}+V_{3}+V_{4}-L=V_{2}+\frac{V_{3}}{2}-1$. From the analysis, we see that one expects $\mathcal O(1)$ answer in $\e$ from the diagrams with either a single quadratic vertex or two cubic vertices and any number of the quartic vertex. A 1PI diagram with only quartic vertices does not contribute to the effective potential for $\phi_{0}$ and can be removed by renormalizing the potential by a constant term.  

We first compute the contribution to the effective potential from diagrams with vertices $(3)$ and $(7)$. A couple of these diagrams are shown in figure~\ref{EffectivePot}. Note that the the other quadratic vertices $(1)$ and $(5)$ are $\mathcal O({\underline n}^{2})$.
\begin{figure}[htpb]
\begin{center}
\vspace{-10 cm}
\includegraphics[width=6in]{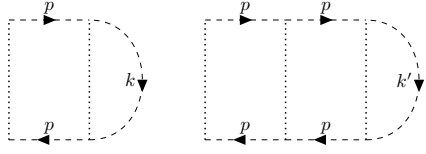}
\vspace{-7 cm}
\caption{Contribution to the effective potential from the vertex $(3)$ and $(7)$.}\label{EffectivePot}
\end{center}
\end{figure}
Summing over all these diagrams, we get
\bea
&&\sum_{n=1}^{\infty}\lambda^{n}\underline {n}\phi_{0}\phi_{0}^{\dagger}\beta\,V_{d}\Big(\int\frac{d^{d}k}{(2\pi)^{d}}\frac{1}{(\frac{\e_{k}}{2}+\delta)^{2}}\Big)^{n-1}\int\frac{d^{d}p}{(2\pi)^{d}}\frac{1}{\frac{\e_{p}}{2}+\delta}\nn\\
&&\qquad\qquad\qquad=\underline {n}\phi_{0}\phi_{0}^{\dagger}\beta\,V_{d}\frac{\lambda\int\frac{d^{d}p}{(2\pi)^{d}}\frac{1}{\frac{\e_{p}}{2}+\delta}}{1-\lambda\int\frac{d^{d}k}{(2\pi)^{d}}\frac{1}{(\frac{\e_{k}}{2}+\delta)^{2}}}=\underline {n}\phi_{0}\phi_{0}^{\dagger}\beta\,V_{d-1}\mathcal C\,.
\eea
Evaluating the above integrals, we obtain
\be
\int\frac{d^{d}p}{(2\pi)^{d}}\frac{1}{\frac{\e_{p}}{2}+\delta}=-\frac{2\delta}{\pi^{2}\e}+\mathcal O(1),\quad \int\frac{d^{d}p}{(2\pi)^{d}}\frac{1}{(\frac{\e_{p}}{2}+\delta)^{2}}=\frac{2}{\pi^{2}\e}+\mathcal O(1)\,.
\ee
Thus, to the leading order in $\e$, we have
\be\label{quadraticplusQuartic2}
\mathcal C=-\delta\frac{\frac{2\lambda}{\pi^{2}\e}}{1-\frac{2\lambda}{\pi^{2}\e}}\,.
\ee
Similarly, we can add all the contributions of the diagram consisting of vertices $(4)$ and $(6)$ together with an arbitrary number of the vertex $(7)$. We find
\be\label{CubiplusQuartic}
\lambda^{2}\phi_{0}^{\dagger}\phi_{0}\underline {n}\beta\,V_{d}\Big(X_{0}+\frac{X_{1}Y_{0}}{1-\lambda\int\frac{d^{d}k}{(2\pi)^{d}}\frac{1}{(\frac{\e_{k}}{2}+\delta)^{2}}}\Big)\,,
\ee
where
\bea
&&X_{0}=\lambda^{2}\int\frac{d^{d}k\,d^{d}p}{(2\pi)^{2d}}\frac{1}{(\frac{\e_{k}}{2}+\delta)(\frac{\e_{p}}{2}+\delta)(\frac{\e_{k+p}}{2}+\delta)},\quad Y_{0}=\lambda\int\frac{d^{d}k}{(2\pi)^{d}}\frac{1}{(\frac{\e_{k}}{2}+\delta)}\,,\nn\\
&&X_{1}=\lambda^{2}\int\frac{d^{d}k\,d^{d}p}{(2\pi)^{2d}}\frac{1}{(\frac{\e_{k}}{2}+\delta)(\frac{\e_{p}}{2}+\delta)^{2}(\frac{\e_{k+p}}{2}+\delta)}\,.
\eea
Evaluating each of these integrals (see appendix~\ref{UsefulIntegration}) to the leading order in $\e$, we get
\be
X_{0}=-\frac{6\lambda^{2}\delta}{\pi^{4}\e^{2}}+\mathcal O(\e),
\quad X_{1}=\frac{2\lambda^{2}}{\pi^{4}\e^{2}}+\mathcal O(\e),\quad Y_{0}=-\frac{2\lambda\delta}{\pi^{2}\e}+\mathcal O(\e)\,.
\ee
Thus, the contribution~\eqref{CubiplusQuartic} to a leading order in $\e$ is
\be\label{CubiplusQuartic2}
-\Big(\frac{2\lambda^{2}\delta}{\pi^{4}\e^{2}}\Big)\frac{3-\frac{4\lambda}{\pi^{2}\e}}{1-\frac{2\lambda}{\pi^{2}\e}}\,.
\ee
Adding the contributions~\eqref{quadraticplusQuartic2} and \eqref{CubiplusQuartic2}, we obtain
\be
-\underline {n}\phi_{0}\phi_{0}^{\dagger}\beta\,V_{d}\frac{2\lambda\delta}{\pi^{2}\e}\times\frac{1+\frac{3\lambda}{\pi^{2}\e}-\frac{4\lambda^{2}}{\pi^{4}\e^{2}}}{1-\frac{2\lambda}{\pi^{2}\e}}\,.
\ee
Note that the above contribution has a Landau pole in $\lambda$. 

Finally, the effective potential becomes
\bea
\bar V(T,\m_{i},\phi_{0})&=&\delta'\,\phi_{0}^{\dagger}\phi_{0}-\int_{\vec p}\Big[E_{p}+2T\ln(1+e^{-\frac{E_{p}}{T}})\Big]+\int_{\vec p} \Big[\frac{1}{2}(\frac{\e_{\vec p}}{2}+\delta\Big)+T\ln\Big(1-e^{-\frac{\frac{\e_{\vec p}}{2}+\delta}{T}}\Big)\Big]\,.\nn\\
\eea
where
\be
\delta'=\delta\Big(1+\frac{2\lambda}{\pi^{2}\e}+\frac{4\lambda^{2}}{\pi^{4}\e^{2}}+\frac{\frac{6\lambda^{2}}{\pi^{4}\e^{2}}}{1-\frac{2\lambda}{\pi^{2}\e}}\Big)\,,
\ee
and
\be
E_{p}^{2}=(\epsilon_{\vec p}-\tilde\m)^{2}+g_{0}^{2}\phi_{0}\phi_{0}^{\dagger}\,.
\ee
In this case, however, we do not encounter any non-trivial minima at $T=0$ as a function of chemical potentials.
\subsection{A different scaling}
In the above analysis, we had a situation where $\phi_0$ and chemical potentials do not scale with $\e$. Now, we consider the case where $\phi_{0}\sim \frac{1}{\sqrt{\e}}$, $\delta\sim\e$ and $\m\sim\e$. In this case, however, the contribution of $\e$ from a given diagram differs from what we have done previously. Taking $\delta\hat\phi_A^\dagger\hat\phi_A$ as an interaction term, we find that a Feynman diagram diverges at most as $\e^{V_\delta-1}$, where the $V_\delta$ is the number of the vertex $\delta\hat\phi_A^\dagger\hat\phi_A$ in the Feynman diagram. The contribution is independent of the number of cubic and quartic vertices. Therefore, for the $\mathcal O(1)$ contributions, we have to consider the diagrams with $V_\delta=1$ and an arbitrary number of cubic (even number) and quartic vertices. As a result, all the monomials of $\phi_0^\dagger\phi_0$ will equally contribute to the effective potential, i.e., summing over all 1PI diagrams, we expect the contribution to the effective potential would be
\be
\delta\sum_{n=1}^\infty C_n(\lambda)(\phi_0^\dagger\phi_0)^n\,.
\ee
This analysis is more complicated, and it would be nice to develop a systematic way to compute all these coefficients. We leave it for the future exercise. 

\section*{Acknowledgments}
We want to thank Shankhadeep Chakrabortty for the useful discussion. We also thank people in the string theory group at TIFR for the valuable conversations and inputs during a seminar where a part of the work was presented. Meenu would like to thank the Council of Scientific and Industrial Research (CSIR), Government of India, for the financial support through a research fellowship (Award No.09/1005(0038)/2020-EMR-I).

\appendix
\section{Useful Integration}\label{UsefulIntegration}
In the main text, we encountered an integral of the following form:
\be
\int\frac{d^{d}p}{(2\pi)^{d}}\frac{d^{d}q}{(2\pi)^{d}}\frac{1}{(p^{2}+b_{3}^{2})(q^{2}+b_{2}^{2})((p-q)^{2}+b_{3}^{2})}\,.
\ee
We are interested in obtaining the leading divergent contribution of the above integral. This can be obtained as follows:
\bea\label{IntegrationResult}
&&\int\frac{d^{d}p}{(2\pi)^{d}}\frac{d^{d}q}{(2\pi)^{d}}\frac{1}{(p^{2}+b_{3}^{2})(q^{2}+b_{2}^{2})((p-q)^{2}+b_{3}^{2})}\nn\\
&&\qquad\qquad=\int\frac{d^{d}p}{(2\pi)^{d}}\frac{d^{d}q}{(2\pi)^{d}}\int_{0}^{1}dx\,dy\,dz\frac{2\delta(x+y+z-1)}{\Big(xp^{2}+yq^{2}+z(p-q)^{2}+b^{2}\Big)^{3}}\nn\\
&&\qquad\qquad=\int\frac{d^{d}p}{(2\pi)^{d}}\frac{d^{d}q}{(2\pi)^{d}}\int_{0}^{1}dx\,dy\,dz\frac{2\delta(x+y+z-1)}{\Big(\alpha p^{2}+\beta q^{2}+b^{2}\Big)^{3}}\nn\\
&&\qquad\qquad=\int\frac{d^{d}p}{(2\pi)^{d}}\frac{d^{d}q}{(2\pi)^{d}}\int_{0}^{1}dx\,dy\,dz\,\delta(x+y+z-1)\int_{0}^{\infty}dt\,t^{2}\,e^{-t(\alpha p^{2}+\beta q^{2}+b^{2})}\nn\\
&&\qquad\qquad=\frac{\Gamma(3-d)}{(4\pi)^{d}}\int_{0}^{1}dx\,dy\,dz\,\delta(x+y+z-1)(\alpha\beta)^{-\frac{d}{2}}(b^{2})^{d-3}\nn\\
&&\qquad\qquad=\frac{\Gamma(3-d)}{(4\pi)^{d}}\int_{x+y\leq1}dx\,dy\frac{1}{(y+x-x^{2}-y^{2}-xy)^{\frac{d}{2}}}\Big(y(b_{2}^{2}-b_{3}^{2})+b_{3}^{2}\Big)^{d-3}\nn\\
&&\qquad\qquad\sim\frac{\Gamma(3-d)}{(4\pi)^{d}}\frac{2}{\e}\Big(2(b_{3}^{2})^{d-3}+(b_{2}^{2})^{d-3}\Big)+\mathcal O(\frac{1}{\e})\,.
\eea
In the above, we have
\be
\alpha=x+z,\quad\beta=y+z-\frac{z^{2}}{x+z}=\frac{xy+yz+zx}{x+z},\quad b^{2}=y(b_{2}^{2}-b_{3}^{2})+b_{3}^{2}\,.
\ee
Thus, it boils down to the integration over the Feynman parameters. Note that the above integral is divergent in $d=4$. The divergence come from the neighbourhood of three points $(x,y)=(0,0),(0,1)$ and $(1,0)$. Near each neighbourhood, the leading divergent is $\frac{2}{\e}$. The result~\eqref{IntegrationResult} is consistent with the exact answer obtained in~\cite{Kleinert2001CriticalPO}.

Another integral we encountered in the main text is
\bea
f(|\phi_{0}|)&=&\int\frac{d^{d}p}{(2\pi)^{d}}\frac{g^{2}|\phi_{0}|}{\sqrt{(\epsilon_{\vec p}-\tilde\m)^{2}+g^{2}|\phi_{0}|^{2}}}\nn\\
&=&\frac{g^{2}|\phi_{0}|\tilde\m^{\frac{d}{2}-1}S_{d-1}}{(2\pi)^{d}}\int_{0}^{\infty}q^{d-1}dq\frac{1}{\sqrt{(\frac{q^{2}}{2}-1)^{2}+\tilde a^{2}}}\nn\\
&=&\frac{g^{2}|\phi_{0}|\tilde\m^{\frac{d}{2}-1}2\pi^{\frac{d}{2}}}{(2\pi)^{d}\Gamma(\frac{d}{2})}\int_{0}^{\infty}q^{d-1}dq\frac{1}{\sqrt{(\frac{q^{2}}{2}-1)^{2}+\tilde a^{2}}}
\eea
Let us evaluate the integral. We have
\bea\label{EffectivePot}
f(|\phi_{0}|)&=&\frac{g^{2}|\phi_{0}|\tilde\m^{\frac{d}{2}-1}2\pi^{\frac{d}{2}}}{(2\pi)^{d}\Gamma(\frac{d}{2})\sqrt{\pi}}\int_{0}^{\infty}q^{d-1}dq\int_{0}^{\infty}dt\,\frac{1}{\sqrt{t}}e^{-t(\frac{q^{2}}{2}-1)^{2}-t\tilde a^{2}}\nn\\
&=&X(\tilde a^{2})\,,
\eea
where
\bea
&&X(\tilde a^{2})=-\frac{g^2 |\phi_{0}| \tilde\mu^{\frac{d}{2}-1}
   \cos \left(\frac{\pi  d}{2}\right)}{2^{d+1}\pi^{\frac{d+1}{2}}\,
   \Gamma(\frac{d}{2})} \Big(2^d  \Gamma
   (1-\frac{d}{2}) \Gamma(\frac{d-1}{2}) \,
   _2F_1\left(\frac{2-d}{4},1-\frac{d}{4};\frac{3-d}{2};\tilde a^2+1\right)\nn\\
   &&\qquad\qquad-2
   \left(\tilde a^2+1\right)^{\frac{d-1}{2}} \Gamma(\frac{1-d}{2}) \Gamma
   (\frac{d}{2}) \,
   _2F_1\left(\frac{d}{4},\frac{d+2}{4};\frac{d+1}{2};\tilde a^2+1\right)\Big)\,.
\eea

\providecommand{\href}[2]{#2}\begingroup\raggedright\endgroup

\end{document}